\documentclass[10pt,journal,compsoc]{IEEEtran}

\def\ps@headings{%
\def\@oddhead{\mbox{}\scriptsize\rightmark \hfil \thepage}%
\def\@evenhead{\scriptsize\thepage \hfil \leftmark\mbox{}}%
\def\@oddfoot{}%
\def\@evenfoot{}}
\usepackage{epsfig, epsf, array, latexsym, graphics, multirow, color}
\usepackage{graphicx}
\usepackage{amsmath}
\usepackage{kpfonts}
\usepackage{cuted}
\usepackage{flushend}
\usepackage[ruled]{algorithm}
\usepackage{algorithmicx}
\usepackage{algpseudocode}

\usepackage{helvet}         % selects Helvetica as sans-serif font
\usepackage{courier}        % selects Courier as typewriter font
\usepackage{type1cm}        % activate if the above 3 fonts are
\usepackage{makeidx}         % allows index generation
\usepackage{graphicx}        % standard LaTeX graphics tool
\usepackage{multicol}        % used for the two-column index
\usepackage[bottom]{footmisc}% places footnotes at page bottom
\newcommand {\mymarginpar}[1]{\marginpar{#1}}
\renewcommand {\marginpar}[1]{} % comment out this command to show labels in the margin

\def\_{\rule{.3em}{.15ex}}      % Get underscore by typing \_.

\newcommand{\ls}[1]
   {\dimen0=\fontdimen6\the\font
    \lineskip=#1\dimen0
    \advance\lineskip.5\fontdimen5\the\font
    \advance\lineskip-\dimen0
    \lineskiplimit=.9\lineskip
    \baselineskip=\lineskip
    \advance\baselineskip\dimen0
    \normallineskip\lineskip
    \normallineskiplimit\lineskiplimit
    \normalbaselineskip\baselineskip
    \ignorespaces
   }
\newcommand {\bearn}{\begin{eqnarray*}}
\newcommand {\eearn}{\end{eqnarray*}}
\newcommand {\barr}{\begin{array}}
\newcommand {\earr}{\end{array}}

\newcommand {\N}{{\cal N}}

\def\defeq{\stackrel{\scriptstyle\rm def}{=}}
\def\twoLineSub#1#2{{#1}\atop{#2}}

\newtheorem{definition}{Definition}
\newtheorem{property}[definition]{Property}
\newtheorem{proposition}[definition]{Proposition}
\newtheorem{lemma}[definition]{Lemma}
\newtheorem{theorem}[definition]{Theorem}
\newtheorem{corollary}[definition]{Corollary}
\newtheorem{example}[definition]{Example}
\newtheorem{remark}[definition]{Remark}
\newtheorem{conjecture}[definition]{Conjecture}
\newtheorem{assumption}[definition]{Assumption}

\newcommand {\benum} {\begin{enumerate}}
\newcommand {\eenum} {\end{enumerate}}

\newcommand {\bdesc} {\begin{description}}
\newcommand {\edesc} {\end{description}}

\newcommand {\bfig}[2] {\begin{figure}[htbp]
                        \centerline {
                         \epsfig{figure={#1},clip=,width={#2}}}}
\newcommand {\brotatefig}[2] {\begin{figure}[htbp]
                        \centerline {
                         \epsfig{figure={#1},clip=,angle=-90,width={#2}}}}
\newcommand {\bfigfirst}[2] {\begin{figure}[h]
                        \centerline {
                        \setlength{\epsfxsize}{#2}
                        \epsffile{#1}}}
\newcommand {\efig}[2]{ \caption{#2}
                        \label{fig:#1}
                        \end{figure}
                        \mymarginpar{fig:#1}}
\newcommand {\erotatefig}[2]{ \caption{#2}
                        \label{fig:#1}
                        \end{figure}
                        \mymarginpar{fig:#1}}

\newcommand {\btab}[1]{
                       \begin{table}
                       \centering
                       \begin{tabular}{#1}}
\newcommand {\etab}[3] {
                       \end{tabular}
                       \caption[#3]{#2}
                       \label{tab:#1}
                       \end{table}
                       \mymarginpar{tab:#1}
                       \vspace{.1in}}

\newcommand {\btabular}[1]{\begin{center}
                       \begin{tabular}{#1}}
\newcommand {\etabular}{\end{tabular}
                       \end{center}}

\newcommand {\bdefin}[1]{\begin{definition}
                      \mymarginpar{def:#1}
                      \label{def:#1} }
\newcommand {\edefin}       {\end{definition}}

\newcommand {\bassum}[1]{\begin{assumption}
                      \mymarginpar{ass:#1}
                      \label{ass:#1} }
\newcommand {\eassum}       {\end{assumption}}

\newcommand {\bpro}[1]{\begin{property}
                      \mymarginpar{pro:#1}
                      \label{pro:#1} }
\newcommand {\epro}   {\end{property}}

\newcommand {\bprop}[1]{\begin{proposition}
                      \mymarginpar{prop:#1}
                      \label{prop:#1} }
\newcommand {\eprop}       {\end{proposition}}
\newcommand {\rprop}[1]{Proposition \ref{prop:#1}}

\newcommand {\blem}[1]{\begin{lemma}
                      \mymarginpar{lem:#1}
                      \label{lem:#1} }
\newcommand {\elem}   {\end{lemma}}

\newcommand {\bthe}[1]{\begin{theorem}
                      \mymarginpar{the:#1}
                      \label{the:#1} }
\newcommand {\ethe}   {\end{theorem}}
\newcommand {\rthe}[1]{Theorem \ref{the:#1}}

\newcommand {\bcor}[1]{\begin{corollary}
                      \mymarginpar{cor:#1}
                      \label{cor:#1} }
\newcommand {\ecor}   {\end{corollary}}

\newcommand {\bax}[1]{\begin{axiom}
                      \mymarginpar{ax:#1}
                      \label{ax:#1} }
\newcommand {\eax}       {\vspace{-.1in} \end{axiom}}

\newcommand {\bconj}[1]{\begin{conjecture}
                      \mymarginpar{conj:#1}
                      \label{conj:#1} }
\newcommand {\econj}       {\vspace{-.1in} \end{conjecture}}

\newcommand {\bex}[2]{\vspace{.1in}
                      \begin{example}
                      \mymarginpar{ex:#1}
                       {\bf #2}
                      \label{ex:#1} \em}
\newcommand {\eex}       {\end{example} \vspace{.3cm} }

\newcommand {\brem}[1]{\begin{remark}
                      \mymarginpar{rem:#1}
                      \label{rem:#1} \em }
\newcommand {\erem}   {\end{remark}}

\newcommand {\beq}[1]{\mymarginpar{eq:#1}
                      \begin{equation}
                      \label{eq:#1} }

\newcommand {\beqno}[1]{\mymarginpar{eq:#1}
                      \begin{eqnarray}
                      \nonumber}

\newcommand {\eeq}       {\end{equation}}
\newcommand {\eeqno}       { && \end{eqnarray}}
\newcommand {\req}[1]{(\ref{eq:#1})}

\newcommand {\bear}[1]{\mymarginpar{eq:#1}
                       \begin{eqnarray}
                       \label{eq:#1} }

\newcommand {\bearno}[1]{\mymarginpar{eq:#1}
                       \begin{eqnarray}
                       \nonumber}

\newcommand {\eear}{\end{eqnarray}}
\newcommand {\eearno}{\end{eqnarray}}
\newcommand {\bsel}{\left \{ \begin{array}{cl}}
\newcommand {\esel}{\end{array} \right.}

\newcommand {\bmat}[1]{\left [ \begin{array}{#1}}
\newcommand {\emat}{\end{array} \right ]}
\newcommand {\bsec}[2]{\mymarginpar{sec:#2}
                       \section{#1}
                       \label{sec:#2} }

\newcommand {\rsec}[1]{Section \ref{sec:#1}}

\newcommand {\bsubsec}[2]{\mymarginpar{sec:#2}
                       \subsection{#1}
                       \label{sec:#2} }

\newcommand {\rsubsec}[1]{Section \ref{sec:#1}}

\def\R{I\kern-0.30em R}
\def\P{I\kern-0.30em P}

\newcommand {\bxfig}[2] {\begin{figure}[htbp]
                        \centerline {
                         \includegraphics[width=#2]{#1}}}
\newcommand {\brotatexfig}[2] {\begin{figure}[htbp]
                        \centerline {
                         \includegraphics[width=#2,angle=90]{#1}}}
\DeclareGraphicsExtensions{.pdf,.jpg,.png}

\def\ex{{\bf\sf E}}
\def\pr{{\bf\sf P}}
\newcommand{\cpr}[2]{{\bf\sf P}(#1\;|\;#2)}

\def\bfmu{{\mbox{\boldmath $\mu$}}}
\def\bfomega{{\mbox{\boldmath $\omega$}}}

\def\bfe{{\mbox{\boldmath $e$}}}

\def\bfg{{\mbox{\boldmath $g$}}}

\def\bfr{{\mbox{\boldmath $r$}}}

\def\bfx{{\mbox{\boldmath $x$}}}
\def\bfy{{\mbox{\boldmath $y$}}}
\def\bfz{{\mbox{\boldmath $z$}}}
\def\bfsx{{\mbox{\boldmath\scriptsize $x$}}}

\def\bfI{{\mbox{\boldmath $I$}}}

\def\bfX{{\mbox{\boldmath $X$}}}
\def\bfY{{\mbox{\boldmath $Y$}}}

\def\bfzero{{\mbox{\boldmath $0$}}}

\def\stge{\ge_{\rm st}}
\def\twoLineSub#1#2{{#1}\atop{#2}}
\def\argmin{\mathop{\rm argmin}}

\def\indicatorFunction#1{{1_{\left\{#1\right\}}}}

\def\intersection{\mathop \bigcap}

\def\ie{{\em i.e.}\ }
\def\pu{P_{URLLC}}
\def\pe{P_{eMBB}}

\begin{document}

\title{Resource Allocation  for URLLC and eMBB Traffic in Uplink Wireless Networks}
\author{
        Duan-Shin~Lee,~\IEEEmembership{Member,~IEEE,}
        Cheng-Shang~Chang,~\IEEEmembership{Fellow,~IEEE}% <-this % stops a space
        Ruhui~Zhang and Mao-Pin Lee 
\IEEEcompsocitemizethanks{\IEEEcompsocthanksitem Duan-Shin~Lee
is with the Department of Computer Science and the Institute of Communications Engineering, 
Cheng-Shang Chang is with the Institute of Communications Engineering,  Ruhui Zhang is with the Department
of Computer Science, and Mao-Pin Lee is  with the Institute of Communications Engineering, 
National Tsing Hua University, Hsinchu 300, Taiwan, R.O.C. (Email:lds@cs.nthu.edu.tw, cschang@ee.nthu.edu.tw,huibrana@gapp.nthu.edu.tw,teddy1998mb@gmail.com.)}
\thanks{This research was supported in part by the Ministry of Science and Technology,
		Taiwan, R.O.C., under Contract 109-2221-E-007-093-MY2.}}

% make the title area
\maketitle

\begin{abstract}
		In this paper we consider two resource allocation problems of URLLC traffic and eMBB 
		traffic in uplink 5G networks.
		We propose to divide frequencies into a common region and a grant-based region.
		Frequencies in the grant-based region can only be used by eMBB traffic, while frequencies
		in the common region can be used by eMBB traffic as well as URLLC traffic.  
		In the first resource allocation problem we propose a two-player game to address the 
		size of the grant-based region and the size of
		the common region.  We show that this game has specific pure Nash equilibria.  
		In the second resource allocation problem we determine the number of packets
		that each eMBB user can transmit in a request-grant cycle.
		We propose a constrained optimization problem to minimize the variance 
		of the number of packets granted to the eMBB users.  We show that a water-filling
		algorithm solves this constrained optimization problem.  From simulation, 
		we show that our scheme, consisting of
		resource allocation according to Nash equilibria of a game, persistent random retransmission
		of URLLC packets and allocation of eMBB packets by a water-filling algorithm, works 
		better than four other heuristic methods.
\end{abstract}
\textbf{keywords}: wireless networks, resource allocation, game theory, constrained optimization,
water-filling algorithm

\bsec{Introduction}{SI}

The fifth-generation networks (5G) and beyond aim to cover
three generic connectivity types: (i) enhanced mobile broadband (eMBB), (ii) ultra-reliable low-latency
communications (URLLC), and (iii) massive machine-type communications (mMTC) (see, e.g.,
\cite{li20175g,bennis2018ultra,Popovski2019} and references therein). The reliability 
defined in 3GPP for supporting URLLC services, such as autonomous driving, drones, and 
augmented/virtual reality, requires a $1-10^{-5}$ success probability of transmitting a layer 2 
packet of length 32 bytes within 1 millisecond. mMTC services are characterized by a large
number of simple devices. Motivated by these emerging needs in 5G, research 
communities in wireless networking
commonly believe that grant-based communications continue to be feasible for eMBB services, but
multiple access schemes in a grant-free manner are more suitable for URLLC and mMTC services.
Many multiple access schemes  have been proposed in the literature recently.  
We refer the readers to \cite{Liu, Yu2021} for references.  In this paper we propose a
method to deliver URLLC packets reliably with their latency bounds, and to schedule
eMBB packets in a fair manner.

Providing satisfactory services to ultra-reliable low-latency communications
(URLLC) traffic and enhanced mobile broadband (eMBB) traffic simultaneously is an interesting and 
challenging research problem.  These two traffic types are characterized by very 
different service requirements.  URLLC traffic demands a latency as low as 1 millisecond
and a packet loss probability as low as $10^{-5}$ \cite{3GPP-1}.  On the other hand, an eMBB user
cares very much about its throughput and also the fairness in the throughput among
all eMBB users.  eMBB services can be efficiently supported by a request and grant paradigm.
However, a grant-based service is not likely to meet the stringent latency requirement of URLLC
services.  A grant-free paradigm is more likely to satisfy the latency requirement of URLLC users.
Non-Orthogonal Multiple Access (NOMA) offers a good solution to the problem above \cite{Khorov2020}.
The whole frequency band can be used for both grant-based eMBB services and grant-free
URLLC services \cite{Gerasin2020, Abreu2019, Abreu2019a, Mollanoori2014, Anand2020}.  Using NOMA
techniques, one can divide the frequency band into a grant-based region and a 
common region in uplink 5G networks.
Using wireless resource blocks in the grant-based region, 
eMBB users make requests, receive grants, and transmit packets according to the grants.  
Properly designed resource allocation algorithms can ensure that eMBB users receive
their services in a fair manner.
Using wireless resource blocks in the common region, URLLC users sporadically transmit
packets in a grant-free manner.  eMBB users also can be granted packets in the common
region.  If so, eMBB users transmit their granted packets at a smaller power in the common 
region.  If eMBB packets collide with URLLC packets in the common region, the eMBB 
packets hopefully can be recovered by the interference cancellation technique \cite{Yu2021}.

In this paper we propose to divide wireless resource blocks into
a common region and a grant-based region. URLLC packets are transmitted sporadically in a 
grant-free manner in the common region.  We propose a persistent retransmission scheme for
the URLLC packets to cope with their stringent latency requirements.  One of the contributions
of this paper is that we analyze the probability that an URLLC packet fails to 
meet its latency requirement.  In this paper we study two resource allocation problems.
In the first problem, we determine the size of the common region 
and the size of the grant-based region.  
We propose to solve this problem by formulating  a two-player game.  The two players of the game 
agent URLLC and agent eMBB, who negotiate resource blocks on behalf of all URLLC users
and eMBB users, respectively. The payoff of the URLLC agent depends on the probability
that an URLLC packet fails to meet its latency requirement.  
The second contribution of this
paper is that we analyze the Nash equilibria of this game.  eMBB users can use resource
blocks in the grant-based region as well as in the common region to transmit their packets.
The transmissions are carried out in a request and grant manner.  
In the second resource allocation problem, we study the allocation of the overall bits that 
can be transmitted in a request-grant cycle to individual eMBB users.  
We solve this problem by formulating a constraint optimization problem to minimize the
variance of the number of bits transmitted by the eMBB users.
The third contribution of this paper is that we show a
water-filling algorithm, solves the nonlinear program.  From simulation, we show 
that our scheme, consisting of
resource allocation according to Nash equilibria of a game, persistent random retransmission
of URLLC packets and allocation of eMBB packets by a water-filling algorithm, works 
better than four other heuristic methods.

The outline of this paper is as follows. In \rsec{frames} we present the frame structure
of the wireless uplink transmission system.
In \rsec{sret-urllc} we present a random persistent retransmission scheme for the URLLC packets.
We present an analysis of the probability that an URLLC packet fails to be transmitted
successfully in this section.
In \rsec{two-player-game} we describe a resource block assignment game to allocate 
the number of resource blocks in the common
region and that in the grant-based region. In \rsec{np} we describe a constrained 
optimization problem of the eMBB users. We present numerical and simulation 
results in \rsec{numerical}. Finally, we conclude this paper in \rsec{conclusions}.

\bsec{Resource blocks and frame structure}{frames}

\begin{figure}[htb]
	\centerline{\includegraphics[width=1.0\linewidth]{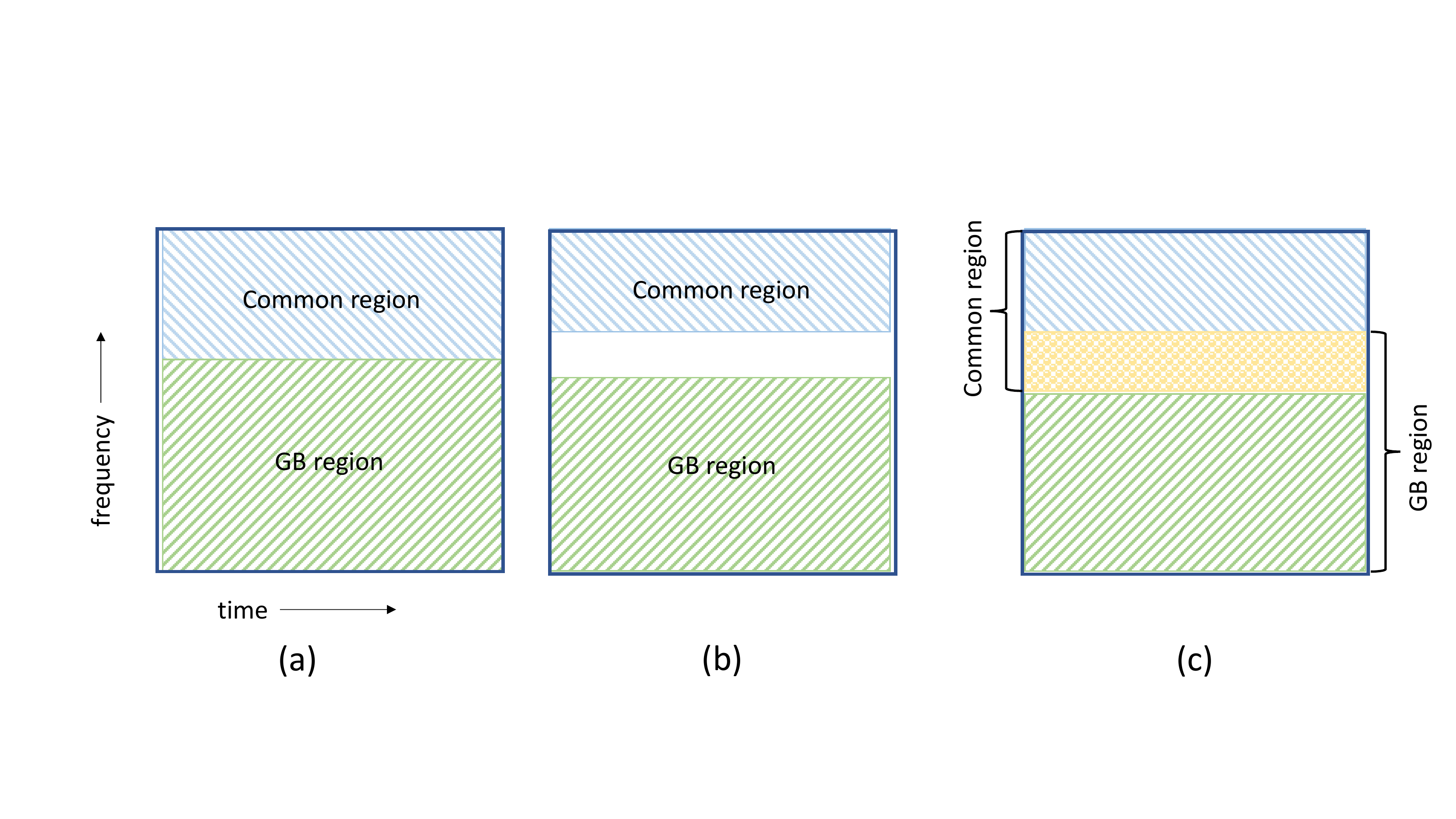}}
	\caption{Resource blocks are divided into a common region and a GB region.  In part (a),
		resource blocks are exactly divided into the two regions.  In part (b), there are resource 
		blocks that are assigned into neither of the two regions.  In part (c), the two regions are 
		overlapped.  In other words, there are resource blocks that belong to the two regions.}
	\label{regions}
\end{figure}

In this paper we propose that the frequency band is divided into a common region
and a grant-based (GB) region in Fig. \ref{regions}.  We assume that URLLC packets can only be
transmitted in the common region in a grant-free manner. The eMBB users make transmission requests.
The eMBB users can use resource blocks in both the grant-based region and the common region to transmit
their granted packets. Since URLLC packets are transmitted in the common region in a grant-free
manner, it is possible that URLLC packets and eMBB packets are collided in the common region.
We assume that eMBB packets are transmitted at a smaller power, so that URLLC packets are
likely to be received successfully.  Then, successive interference cancellation techniques
are applied to recover the collided eMBB packets.

\begin{figure}[htb]
	\centerline{\includegraphics[width=1.0\linewidth]{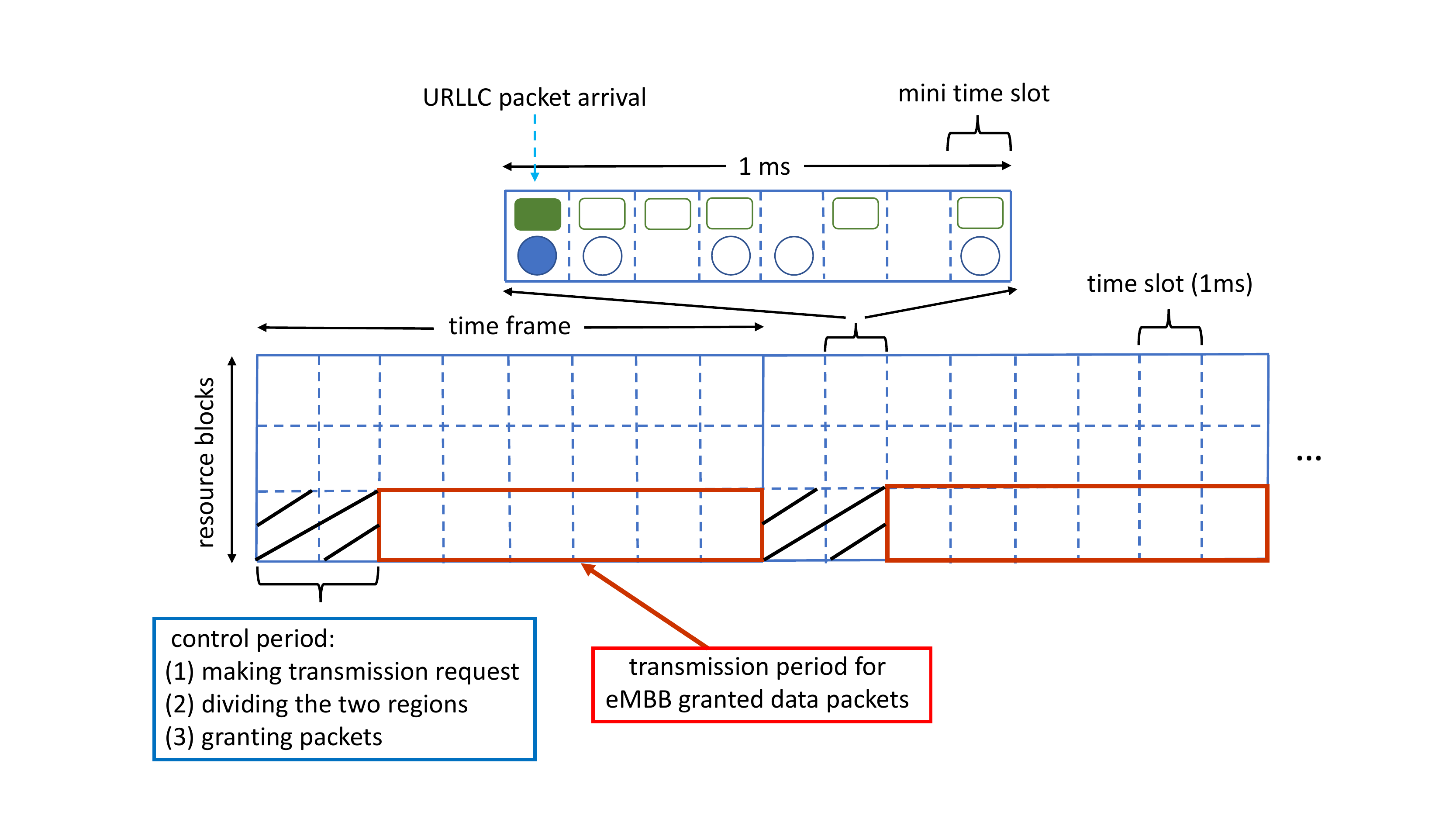}}
	\caption{Schematic description of time slots, mini time slots, time frames and resource blocks. Two URLLC users may transmit packets in the
		same mini time slot. The two packets are shown in rectangles and circles,
		respectively. The two packets collide and are retransmitted persistently
		and randomly.}
	\label{sys}
\end{figure}

In this paper we assume that time is discrete and is divided into time slots.  Multiple
time slots are grouped into a time frame.  A time frame is divided 
into two periods, including a control period 
and a transmission period.  In control periods, 
eMBB users make transmission requests. In transmission periods, eMBB users 
transmit their data packets that are granted by a base station.  In the control 
period, the base station also announces how
resource blocks are divided into a common region and a grant-based region. 
In each time slot, one eMBB packet can be transmitted.
It has been widely recognized that conventional automatic repeat request (ARQ) mechanisms such
as acknowledgments or timeouts cannot solve the reliability issue of URLLC traffic due
to its stringent delay requirement.  To cope with the ultra stringent latency
requirement, it has been proposed that time slots are divided into mini time slots
\cite{Elayoubi2019, 3GPP, Wang2017}.  An URLLC packet can be transmitted in one mini time slot.
We assume that a time slot is divided into $\tau$ mini time slots.  A typical length
of a time slot is 1 millisecond (ms), and a typical value for $\tau$ is $8$.
We show time frames, time slots and mini time slots in Fig. \ref{sys}.

We summarize in Table \ref{symbols} some symbols used in this paper.

%\begin{table}[!htb]
\begin{table}
	\renewcommand\arraystretch{1.2}
	\setlength{\tabcolsep}{10pt}
	\begin{center}
		\begin{tabular}{| p{1 cm} | p{5.5cm} |}\hline
			\textbf{Symbol} & \textbf{Definition}  \\\hline
			$\rho$ &  URLLC packet arrival rate \\\hline
			$\tau$ &  Delay requirement of an URLLC packet \\\hline
			$p$ &  URLLC packet retransmission probability  \\\hline
			$N$ &  Total number of resource blocks \\\hline
			$\epsilon$ & Upper bound of the URLLC packet loss probability  \\\hline
			$E_k$ & Event that a tagged URLLC packet cannot be successfully transmitted given that the common region has $k$ resource blocks  \\\hline
			$b$ & Cost of resource blocks \\\hline	
			$m$ & Number of eMBB users\\\hline	
			$c$ & Number of bits that the eMBB users can transmit using one resource block in 
			the grant-based region\\\hline
		\end{tabular}
	\end{center}
	\caption{Definition of parameters used in this paper.}
	\label{symbols}
\end{table}

\bsec{Retransmission scheme for URLLC traffic}{sret-urllc}

URLLC traffic is characterized by its ultra stringent delay requirement and extremely high
reliability requirement.  An URLLC packet needs to be delivered to its destination within
1 millisecond with a success probability higher than $1-10^{-5}$.  
The ultra low latency requirement makes grant-based transmission mode infeasible 
for URLLC services.  It is commonly believed that grant-free transmission
is suitable for URLLC services \cite{Jacobsen2017}.  

The traditional semi-persistent scheduling (SPS) scheme \cite{3GPP-1}
pre-schedules resources before sessions begin.  
SPS would be suitable for URLLC traffic if the traffic were periodic and predictable.  
However, URLLC traffic is expected to be sporadic.  SPS could be very 
inefficient in resource utilization.  
It has been widely recognized that conventional automatic repeat request (ARQ) mechanisms such
as acknowledgments or timeouts cannot solve the reliability issue of URLLC traffic due
to its stringent delay requirement \cite{Elayoubi2019, 3GPP, Wang2017}.  Packets must be 
retransmitted persistently without waiting for their negative acknowledgments 
or expiration of timeout clocks.  That is, each URLLC packet will be transmitted 
multiple number of times.  An URLLC packet is successfully received if at least one
copy is received successfully. 
\iffalse Specifically, suppose that the delay requirement of an URLLC packet corresponds to
$\tau$ mini time slots.  After an URLLC user transmits its packet, it retransmits the
same packet $\tau-1$ times in the subsequent mini time slots using the same resource block. \fi
This approach has been adopted by the 3GPP standard \cite{3GPP}.  
However, the multiple access nature of the uplink communication makes it possible that
multiple URLLC users attempt to transmit their packets using the same resource block
in the same mini time slot.  A collision thus can happen.  A persistent transmission scheme
would imply that all subsequent retransmissions are also collided.  
In this paper we propose a randomized persistent transmission scheme for 
URLLC traffic.  Suppose that up to $\tau$ copies of each URLLC packet are transmitted.
The first copy is transmitted at the time when it arrives.
Instead of repeating the transmission of the remaining $\tau-1$ copies, one in each of the 
$\tau-1$ subsequent mini time slots, each copy is transmitted with probability $p$.
With probability $1-p$, the URLLC user does not transmit a copy.  This mechanism can
reduce the number of subsequent collisions and let the collided URLLC users have
a chance to successfully transmit their packets.   Fig. \ref{sys}
contains a graphical illustration. A time slot is divided into several 
mini time slots. On the upper side of Fig. \ref{sys}, the
two URLLC users attempt to transmit packets. Original packet
transmissions are shown by colored shapes, and retransmissions are shown by
uncolored shapes. Both of the two URLLC users transmit in the first
mini time slot of a specific time slot using the first resource block.  
They persistently and randomly retransmit.
The first user (shown in rectangle) successfully retransmits its packet 
in the second retransmission attempt.
The second user (shown in circle) successfully retransmits its packet 
in the third retransmission attempt.

We now present an analysis of the probability that a tagged URLLC packet cannot
be successfully transmitted.  Suppose that there are $n_1$ resource blocks in the common
region.  Assume that there are $n$ URLLC users.  Assume that URLLC packets 
arrive according to a Poisson process with rate $\rho$ packets per mini
time slot.  Assume that each URLLC user randomly chooses a resource block 
to transmit its packet. It follows that the URLLC packet arrivals to a 
specific resource block, say block $j$, are Poisson with rate $\rho/n_1$ packets 
per mini time slot.  Let $\tilde\rho=\rho/n_1$ be the traffic intensity 
per resource block.  We consider a specific packet
called tagged packet.  We label the mini time slot that
the tagged URLLC packet arrives as mini time slot $0$.
The sender of this tagged packet continuously attempts to retransmit the packet with
probability $p$ in each of the next $\tau-1$ mini time slots.  The sender does not retransmit
the packet with probability $1-p$ in each of the next $\tau-1$ mini time slots.   
Let $E_{n_1}$ denote the event that the tagged URLLC packet cannot be successfully
transmitted given that the common region has $n_1$ resource blocks.  

We now analyze $\pr(E_{n_1})$. Recall that we denote the URLLC packet
arrival rate to a resource block by $\tilde\rho=\rho/n_1$.  We also simply denote
the event that the tagged packet fails to be successfully transmitted by $E$, rather than
$E_{n_1}$.  Let $G_i$ be the event that the tagged packet is not 
successfully transmitted in mini
time slot $i$ for $i=0, 1, \ldots, \tau-1$.  Then,
\[
E=G_0\cap G_1\cap \cdots \cap G_{\tau-1}.
\]
Let $X_j$ be the number of arrivals in mini time slot $j$ not including the tagged packet
for $j=-(\tau-1), -(\tau-1)+1, \ldots, -1, 0, 1, \ldots, \tau-1$.  Let $A$ denote
the event that $X_j=x_j$ for $-(\tau-1)\le j\le \tau-1$.  That is,
\[
A=\{X_j = x_j, -(\tau-1)\le j\le \tau-1\}.
\] 
Conditioning on event $A$, 
events $G_0, G_1, \ldots, G_{\tau-1}$ are
independent. By the law of total probability and conditional independence, one has
\begin{align}
\pr(E) &= \sum_{\twoLineSub{x_j=0}{-(\tau-1)\le j\le \tau-1}}^\infty 
\pr\left(\intersection_{i=0}^{\tau-1} G_i
\;\Biggl|\; A\right)\cdot \pr(A) \nonumber\\
&= \sum_{\twoLineSub{x_j=0}{-(\tau-1)\le j\le \tau-1}}^\infty \prod_{i=0}^{\tau-1} 
\cpr{G_i}{A}\cdot \pr(A),\label{total-prob}
\end{align} 
where
\begin{align}
\pr(A) &= \pr(X_j=x_j, -(\tau-1)\le j\le \tau-1) \nonumber\\
&= \prod_{j=-(\tau-1)}^{\tau-1} \frac{e^{-\tilde\rho}\tilde\rho^{x_j}}{x_j !}.
\end{align}
Now we derive the conditional probabilities in (\ref{total-prob}).  
If there are new arrivals in mini time slot $0$, the tagged packet cannot be 
successfully transmitted.  Assume that there are no
new arrivals in mini time slot $0$, the tagged packet still cannot be successfully 
transmitted if there is at least one arrival in mini time slots $-(\tau-1), 
-(\tau-1)+1, \ldots, -1$ that is
retransmitted in mini time slot $0$.  Combining these possibilities, we have
\begin{align}
	%\cpr{G_0}{X_j=x_j, -(\tau-1)\le j\le \tau-1} \nonumber\\
	\cpr{G_0}{A}&
	= \indicatorFunction{x_0\ge 1}
	+\indicatorFunction{x_0=0}\left(1-(1-p)^{\sum_{i=-(\tau-1)}^{-1} x_i}\right) \nonumber\\
	&=1-\indicatorFunction{x_0=0}(1-p)^{\sum_{i=-(\tau-1)}^{-1} x_i},\label{G_0}
\end{align}
where notation $\indicatorFunction{A}$ is an indicator function
of event $A$.  The value of $\indicatorFunction{A}$ is equal to $1$ if event $A$ 
is true, and is equal to $0$ otherwise.
For $i=1, 2, \ldots, \tau-1$, we have
\begin{align}
%	&\cpr{G_i}{X_j=x_j, -(\tau-1)\le j\le \tau-1} \nonumber\\
&\cpr{G_i}{A}\nonumber\\
	&=(1-p)\nonumber\\
     &+p\cdot\left[\indicatorFunction{x_i\ge 1}+\indicatorFunction{x_i=0} 
	\left(1-(1-p)^{\sum_{j=i-(\tau-1)}^{i-1} x_j}\right) \right]\nonumber\\
	&=1-p\cdot \indicatorFunction{x_i=0}(1-p)^{\sum_{j=i-(\tau-1)}^{i-1} 
		x_j}.\label{G_i}
\end{align}
In (\ref{G_0}) and (\ref{G_i}), notation $\indicatorFunction{A}$ is an indicator function
of event $A$.  The value of $\indicatorFunction{A}$ is equal to $1$ if event $A$ is true, and is equal to $0$ otherwise.
Substituting (\ref{G_0}) and (\ref{G_i}) into (\ref{total-prob}), we obtain
an expression for $\pr(E)$.  For general values of $\tau$, we are not able
to obtain a closed form expression for $\pr(E)$.  However, we show in the next 
proposition that $\pr(E)$ is an increasing function of $\tilde\rho$.  This result
is used to study the Nash equilibria of a resource allocation game in \rsubsec{two-player-game}.
The proof is presented in Appendix A. 
\bprop{incr}
For any $p$ and $\tau$, $\pr(E)$ is increasing with respect to $\tilde\rho$.
\eprop

To proceed further, we consider a specific case, in which $\tau=3$.  
That is, each new arrival has two opportunities
to be retransmitted.   Substituting Eqs. (\ref{G_0}) and (\ref{G_i}) into 
(\ref{total-prob}), we have
\begin{align}
&\cpr{G_0}{A}\cpr{G_1}{A}\cpr{G_2}{A}  \nonumber\\
	&\quad =\left(1-\indicatorFunction{x_0=0}(1-p)^{x_{-2}+x_{-1}}\right)\nonumber\\
	&\quad  \cdot\left(1-p\cdot\indicatorFunction{x_1=0}(1-p)^{x_{-1}+x_{0}}\right) \nonumber\\
	&\quad  \cdot\left(1-p\cdot\indicatorFunction{x_2=0}(1-p)^{x_{0}+x_{1}}\right).\label{tau=3}
\end{align}
In addition, from the independent increment property of Poisson process, we have
\beq{indep-poisson}
\pr(A)=\prod_{j=-2}^{2} \frac{e^{-\tilde\rho}\tilde\rho^{x_j}}{x_j !}.
\eeq
Substituting (\ref{tau=3}) and \req{indep-poisson} into (\ref{total-prob}) and simplifying, we obtain
\begin{align}
	\pr(E) &= 1-(1+2p)\cdot\exp(-3\tilde\rho+2(1-p)\tilde\rho)\nonumber\\
	&\quad+p(1+p)\cdot\exp(-4\tilde\rho+(1-p)\tilde\rho+(1-p)^2 \tilde\rho) \nonumber\\
	&\quad+p\cdot\exp(-5\tilde\rho+3(1-p)\tilde\rho)\nonumber\nonumber\\
	&\quad-p^2\cdot\exp(-5\tilde\rho+(1-p)\tilde\rho+(1-p)^2\tilde\rho).\label{final-tau=3}
\end{align}
A Taylor expansion of (\ref{final-tau=3}) around $\tilde\rho=0$ is
\begin{align}
	\pr(E) &\approx (1-p^2+p^3)\tilde\rho+o(\tilde\rho^2).
	%\frac{-1+2p+7p^2-4p^4+p^5}{2}\tilde\rho^2 +o(\tilde\rho^3).
	\label{Taylor}
\end{align}

For general values of $\tau$, it is difficult to evaluate 
(\ref{total-prob}).  Since URLLC traffic requires a very low packet
loss probability of the order $10^{-5}$, the system must operate in a light 
traffic condition.  We now propose a light traffic approximation 
for $\pr(E)$ with general values of $\tau$.  It is interesting to
point out that the order of this light traffic approximation with $\tau=3$ 
agrees with the Taylor expansion in (\ref{Taylor}).

We consider a light traffic model, in which excluding the tagged 
packet in mini time slot $0$, in mini time slots in range $[-(\tau-1), \tau-1]$, 
there is either exactly one packet arrival, 
or there are no packet arrivals.  With probability $1-(2\tau-1)\tilde\rho$
there are no additional arrivals.  With probability $(2\tau-1)\tilde\rho$, 
there is exactly one arrival.  If there is exactly one arrival, this packet 
arrives in any one of the $2\tau-1$ mini time slots in
range $[-(\tau-1), \tau-1]$ with an equal probability.  
We now derive $\pr(E)$ for this light traffic model.  Let $H_i$ be the event that
a packet arrives in mini time slot $i$, where $-(\tau-1)\le i\le \tau-1$.
If $1\le i\le \tau-1$,
\beq{lt-1}
\cpr{E}{H_i}=0.
\eeq
If $-(\tau-1)\le i\le -1$,
\beq{lt-2}
\cpr{E}{H_i}=p (1-p+p\cdot p)^{\tau+i-1}\cdot (1-p)^{-i}.
\eeq
If $i=0$,
\beq{lt-3}
\cpr{E}{H_0}=(1-p+p\cdot p)^{\tau-1}.
\eeq
For $i$ in range $[-(\tau-1), \tau-1]$,
\beq{lt-4}
\pr(H_i)=\frac{(2\tau-1)\tilde\rho}{2\tau-1}=\tilde\rho.
\eeq
Let $H^c$ denote the event that there are no additional packet arrivals.  Then,
\begin{align}
	\cpr{E}{H^c} &= 0 \nonumber\\
	\pr(H^c) &= 1-(2\tau-1)\tilde\rho.\label{lt-5}
\end{align}
By the law of total probability and Eqs. \req{lt-1} - (\ref{lt-5}), we have
\begin{align}
	&\pr(E) \nonumber\\
	&= \sum_{i=-(\tau-1)}^{-1} \cpr{E}{H_i}\pr(H_i)\nonumber\\
      & +\cpr{E}{H_0} \pr(H_0)+\cpr{E}{H^c}\pr(H^c)\nonumber\\   
	&=\Bigg(\sum_{i=-(\tau-1)}^{-1} p(1-p+p^2)^{\tau+i-1}(1-p)^{-i} \nonumber\\
	& +(1-p+p^2)^{\tau-1}\Bigg) \tilde\rho. \label{lta}
\end{align} 
In a special case where $\tau=3$, (\ref{lta}) agrees with (\ref{Taylor}).

\bsec{A two-player game}{two-player-game}

In this section we propose a two-step procedure to the resource block allocation to
URLLC users and eMBB users.  We first apply the Nash equilibria of a two-player game
to determine the number of resource blocks in the common region and that in the grant-based
region. We then apply the solution of a constrained nonlinear optimization
problem to allocate the number of bits granted to the eMBB users.  
We explain the two-player game in this section.  We present the constrained nonlinear
optimization problem in \rsec{np}.

We now explain a two-player resource allocation game.  The two players are an URLLC agent 
and an eMBB agent, negotiating resource blocks on behalf of URLLC users and eMBB users, respectively. 
The URLLC agent negotiates the number of resource blocks in the common region. The eMBB agent negotiates the number of resource blocks in the grant-based
region.
Assume that there are $n$ URLLC users and $m$ eMBB users.  Assume that
there are totally $N$ resource blocks to be allocated.  The set of actions for the URLLC agent
is $\{0, 1, \ldots, N\}$.  The eMBB agent has the same set of actions.
When the URLLC agent takes action $n_1$ and the eMBB agent takes action $n_2$, where
$0\le n_1, n_2\le N$, the payoff to the URLLC agent is denoted by 
$P_{URLLC}(n_1,n_2)$ and the payoff
to the eMBB agent is denoted by $P_{eMBB}(n_1,n_2)$. Recall that
$E_k$ denotes the event that a tagged URLLC packet cannot be successfully
transmitted given that the common region has $k$ resource blocks.
From the light traffic approximation of (\ref{lta}), we have
\begin{align}
	&\pr(E_k)=\nonumber\\
	&\quad\left\{\begin{array}{ll}
		\Bigg(\displaystyle\sum_{i=-(\tau-1)}^{-1} p(1-p+p^2)^{\tau+i-1}(1-p)^{-i}\\
		+(1-p+p^2)^{\tau-1}\Bigg)
           \cdot \rho/k  &\mbox{if $k\geq1$}\\
		1 &\mbox{if $k=0$.}\end{array}\right.\label{pek}		
\end{align} 
We assume that the URLLC agent's payoff is
\beq{payoff-1}
\pu(n_1,n_2)=\indicatorFunction{\pr(E_{\min(n_1,N-n_2)})\le \epsilon}
\cdot\indicatorFunction{n_1+n_2\leq N}-\frac{n_1}{N}\cdot b,
\eeq  
where $\epsilon$ is an upper bound of the URLLC packet loss probability, 
and $b$ is a cost of using $N$ resource blocks.  We assume that
\beq{b}
0< b<1.
\eeq
Suppose that an eMBB user can transmit $c$ bits using one resource block in
the grant-based region in a time frame.
If action profile $(n_1,n_2)$ is taken by the two players, the grant-based region has
$\min(n_2, N-n_1)$ resource blocks that are not overlapped with resource blocks in
the common region.  Thus, the eMBB users can transmit $\min(n_2, N-n_1)\cdot c$ bits in
this region.  On the other hand, the common region has $\min(n_1, N-n_2)$ resource blocks
that are not overlapped with resource blocks in the grant-based region.  
We assume that the eMBB users transmit
at a smaller power, and their packets are recovered by interference cancellation if
they are collided with URLLC packets.  Thus, the number of bits that the eMBB users
can transmit in this region is $a\cdot\min(n_1, N-n_2)\cdot c$, where $0<a<1$.
Parameter $a$ reflects the fact that an eMBB user can
transmit less data using resource blocks in the common region.  
We refer the readers to Fig. \ref{regions} for an illustration
of these regions.  The total number of bits that eMBB users can 
potentially transmit in a time frame given that action profile $(n_1,n_2)$ is taken is
$T_{n_1, n_2}$, where
\begin{align}
 T_{n_1, n_2}=[\min(n_2,N-n_1)+a \cdot \min(n_1,N-n_2)]\cdot c. \label{trr}
\end{align}
Define
\begin{align}
	&n_2^*(n_1) =\nonumber\\
	&\quad\left\{\begin{array}{ll}
		\min\{n_2 :T_{n_1,n_2}\ge r\} & \mbox{if $\max_{0\leq n_2\leq N} T_{n_1,n_2}\ge r$}\\
		N & \mbox{otherwise,}
	\end{array}\right.\label{jst}
\end{align}
where $r$ is the number of bits requested by eMBB users collectively in a time frame.
Quantity $n_2^*(n_1)$ is the minimum number of resource blocks for the common region
in order for eMBB users to transmit $r$ bits.
We assume that the eMBB agent's payoff is
\begin{align}
    &\pe(n_1,n_2)= \nonumber\\
    &\quad\left\{\begin{array}{ll}
    	\frac{n_2^*(n_1)}{N}\cdot\indicatorFunction{n_1+n_2\leq N}-\frac{n_2}{N}\cdot b & \mbox{if $n_2> n_2^*(n_1)$}\\
    	\frac{n_2}{N}\cdot\indicatorFunction{n_1+n_2\leq N}-\frac{n_2}{N}\cdot b & 
    	\mbox{if $n_2 \leq n_2^*(n_1)$.}
      \end{array}\right.\label{payoff-2}
\end{align}

If $\min_{0\le n_1\le N} \pr(E_{n_1})\le \epsilon$, define
%\beq{istar}
%i^* = \min \{i : \pr(E_i)\le\epsilon\}.
%\eeq
\beq{istar}
n_1^* = \min\{n_1: \pr(E_{n_1})\le \epsilon\}.
\eeq
Clearly, 
$1\le n_1^*\le N$.

The following theorem characterizes the Nash equilibria of the two-player
resource allocation game.  Its proof is presented in Appendix B.
\bthe{Nash}
If $\min_{0\le i\le N} \pr(E_i)\le \epsilon$, there are three cases as follows.
\begin{enumerate}
	\item If 
	\beq{case1}
	n_2^*(0) \leq N-n_1^*\ \mbox{and}\ n_2^*(n_1^*) \leq N-n_1^*.
	\eeq
	action profile $(n_1^*, n_2^*(n_1^*))$ 
	is the only one pure Nash equilibrium of the two-player resource allocation game.
	\item If 
	\beq{case2}
	n_2^*(0) > N-n_1^*\ \mbox{and}\ n_2^*(n_1^*) \leq N-n_1^*.
	\eeq
	holds, action profiles $(n_1^*, n_2^*(n_1^*))$ 
	and $(0, n_2^*(0))$ are the only pure Nash equilibria of the two-player resource
	      allocation game. In addition,
	      \begin{enumerate}
	      	  \item if $n_1^*<N$ or $n_2^*(0)-n_2^*(n_1^*)<N$, $(n_1^*, n_2^*(n_1^*))$ is socially better
	      	  than the other equilibrium.
	      	  \item If  $n_1^*=N$ and $n_2^*(0)-n_2^*(n_1^*)=N$, both $(n_1^*, n_2^*(n_1^*))$ and $(0, n_2^*(0))$ 
	      	  are socially optimal.
	      \end{enumerate}
	\item If 
	\beq{case3}
	n_2^*(0) > N-n_1^*\ \mbox{and}\ n_2^*(n_1^*) > N-n_1^*.
	\eeq
	holds, action profiles $(n_1^*, N-n_1^*)$ 
	and $(0, n_2^*(0))$ are the only pure Nash equilibria of the two-player resource
	      allocation game.  In addition, 
	      \begin{enumerate}
	      	  \item if $n_1^*<N$ or $n_2^*(0)<N$, $(n_1^*, N-n_1^*)$ is socially better than the
	      	  other equilibrium.
	      	  \item If $n_1^*=N$ and $n_2^*(0)=N$, both $(n_1^*, N-n_1^*)$ and $(0, n_2^*(0))$ are 
	      	  socially optimal.
	      \end{enumerate} 
\end{enumerate}
If $\min_{0\le i\le N} \pr(E_i)> \epsilon$, action profile $(0, n_2^*(0))$ is 
the only one pure Nash equilibrium of the two-player resource allocation game.
\ethe

{\em Remark.} We remark that \rthe{Nash} contains only three cases (\ie \req{case1},
\req{case2} and \req{case3}) when 
$\min_{0\le i\le N} \pr(E_i)\le \epsilon$.  We now show that 
$n_2^*(0) \leq N-n_1^*$ and $n_2^*(n_1^*) \leq N-n_1^*$ can not occur
at the same time.  To see this, note from the definition of $n_1^*$ in \req{istar}
that
\begin{align}
	T_{n_1^*,N-n_1^*} &=[\min(N-n_1^*,N-n_1^*)+a \cdot \min(n_1^*,n_1^*)]\cdot c \nonumber\\
	&=(N-n_1^*+a \cdot n_1^*)\cdot c \nonumber\\
	&>[\min(N-n_1^*,N)+a \cdot \min(0,n_1^*)]\cdot c\nonumber\\
	&=T_{0,N-n_1^*}.\label{ineq}
\end{align}
Suppose that $n_2^*(0) \leq N-n_1^*$.
From (\ref{ineq}), we have $T_{n_1^*,N-n_1^*}>T_{0,N-n_1^*} \geq r$,
Thus, $n_2^*(n_1^*) \leq N-n_1^*$.

\bsec{A constrained optimization problem}{np}

In this section we present a constrained nonlinear program to assign the 
number of bits to be granted to the eMBB users. As before, we assume that there are $m$ eMBB
users.  eMBB user $j$, where $1\le j\le m$, has cumulatively been granted $z_j(t)$ bits up to time
frame $t$.  Also assume that user $j$ requests to transmit $r_j$ bits in time frame $t+1$.
We assume that $r_j>0$ for all $1\le j\le m$.
Suppose that user $j$ is granted to transmit $x_j$ bits in time frame $t+1$.
It follows that
\beq{recrusive-z}
z_j(t+1)=z_j(t)+x_j.
\eeq
We aim to find $\{x_1, x_2, \ldots, x_m\}$ such that the sample variance of $\{z_1(t+1),
z_2(t+1),\ldots, z_m(t+1)\}$ is minimized.  That is, we attempt to minimize 
\beq{var}
\frac{1}{m-1}\sum_{j=1}^m \left(z_j(t+1)-\frac{1}{m}\sum_{k=1}^m z_k(t+1)\right)^2. 
\eeq
Since the units of $\{r_j, 1\le j\le m\}$, $\{z_j, 1\le j\le m\}$, and
$\{x_j, 1\le j\le m\}$ are bits, they tend to be large numbers.  It is reasonable to
ignore their integer constraints and approximate them by real numbers.  

Suppose that the common region has $n_1$ resource blocks and the grant-based region
has $n_2$ resource blocks.  We propose to use the two-player game presented in
\rsec{two-player-game} to determine $n_1$ and $n_2$.  Specifically, we assume
that $(n_1, n_2)$ is the only one pure strategy Nash equilibrium if the game has only 
one pure strategy Nash equilibrium, and it is the socially optimal Nash 
equilibrium otherwise. In both cases, $n_1+n_2 \le N$.  According to (\ref{trr}), at most
\begin{align}
	L=\min\left((n_2+a\cdot n_1)\cdot c, \sum_{k=1}^m r_k\right)\label{ll}
\end{align}
eMBB bits can be transmitted in time frame $t+1$, where $r_k$ is the number of
bits that user $k$ requests to transmit.
We rewrite \req{var} using $L$, \ie
\begin{align}
  &z_j(t+1)-\frac{1}{m}\sum_{k=1}^m z_k(t+1) \nonumber\\
    &= z_j(t)+x_j- \frac{1}{m}
	\left(\sum_{k=1}^m z_k(t)+\sum_{k=1}^m x_k\right) \nonumber\\
	&= z_j(t)+x_j- \frac{1}{m}\left(\sum_{k=1}^m z_k(t)+L\right) \nonumber\\
	&= x_j-\eta_j,\label{objfunc}
\end{align}
where 
\beq{eta_j}
\eta_j=\frac{1}{m}\left(\sum_{k=1}^m z_k(t) +L\right)-z_j(t)
\eeq
and we have set
\[
\sum_{k=1}^m x_k =L.
\]
For notational simplicity, we shall use notation $z_j$  without $t$ in the rest of this paper.
Define 
\[
F_j(x)=\frac{(x-\eta_j)^2}{m-1}.
\]
We have reached the following separable convex nonlinear program with linear constraints.
\beq{cnp}
	\min_{x_1, x_2, \ldots, x_m}  \sum_{j=1}^m F_j(x_j)
	\eeq
subject to
\begin{align}
	&\sum_{j=1}^m x_j =L \label{eq-constraint}\\
	&0\le x_j\le r_j \ \  \mbox{for all $j$}.\label{ineq-constraint}
\end{align}
We use bold face letters to denote vectors. Let $\bfx=(x_1, x_2, \ldots, \\
x_m)$, $\bfr=(r_1, r_2, \ldots, r_m)$, $\bfe=(1, 1, \ldots, 1)$ and $\bfzero=(0, 0, \ldots, 0)$.  
Expressing \req{cnp}, (\ref{eq-constraint}) and
(\ref{ineq-constraint}) in terms of vectors, we have
\beq{vcnp}
\min_{\bfsx}  F(\bfx) 
\eeq
subject to
\begin{align}
	h(\bfx) &=0 \label{L-constraint}\\
	\bfg_1(\bfx) &\le \bfzero \label{ineq-constraint-1}\\
	\bfg_2(\bfx) &\le \bfzero, \label{ineq-constraint-2}
\end{align}
where 
\begin{align}
	F(\bfx) &\defeq \sum_{i=1}^m F_i(x_i) \label{def-F}\\
	h(\bfx) &\defeq \sum_{j=1}^m x_j -L \label{def-h}\\
	\bfg_1(\bfx) &\defeq -\bfx \label{def-g1}\\
	\bfg_2(\bfx) &\defeq \bfx-\bfr. \label{def-g2}
\end{align}
Suppose that $\bfx^*$ is a regular point that satisfies constraints (\ref{L-constraint}),
(\ref{ineq-constraint-1}) and (\ref{ineq-constraint-2}).  Then by 
the Karush-Kuhn-Tucker theorem \cite[p. 352]{LueD.84}, a necessary condition 
for $\bfx^*$ to be a relative minimum point
for the problem in \req{vcnp} is that there exist a real number $\lambda$ 
and column vectors $\bfmu$ and $\bfomega$ in $R^m$ with $\bfmu\ge \bfzero$ and 
$\bfomega\ge \bfzero$ such that
\begin{align}
\nabla F(\bfx^*)+\lambda \nabla h(\bfx^*)+\bfmu^T \nabla \bfg_1(\bfx^*)
+\bfomega^T \nabla \bfg_2(\bfx^*) &=\bfzero\label{KKT} \\
\bfmu^T \bfg_1(\bfx^*)&=\bfzero \label{complementary-1}\\
\bfomega^T \bfg_2(\bfx^*)&=\bfzero, \label{complementary-2}
\end{align}
where $\bfmu^T$ denotes the transpose of $\bfmu$.  We remark that $\nabla F(\bfx)$
and $\nabla h(\bfx)$ are $1\times m$ row vectors.  $\nabla \bfg_i(\bfx^*)$, $i=1, 2$,
are $m\times m$ matrices.  
An inequality constraint such as 
(\ref{ineq-constraint-1}) is said to be active at a 
feasible point $\bfx$ if $(\bfg_1)_i(\bfx)=0$ for some $1\le i\le m$.  
Otherwise, the inequality constraint is said to be inactive.  An entry of $\bfmu$ or
$\bfomega$ may be non-zero only if the corresponding constraint is active.  
If one knew in advance which constraints were active, the solution of (\ref{KKT}) can
easily be solved.  Specifically, let sets $S$, $S_1$ and $S_2$ be defined as
\begin{align*}
	S &=\{j: \mbox{(\ref{ineq-constraint-1}) and (\ref{ineq-constraint-2}) are both
		inactive}\} \\
	S_1 &=\{j: \mbox{(\ref{ineq-constraint-1}) is active and (\ref{ineq-constraint-2}) is
		inactive}\} \\
	S_2 &=\{j: \mbox{(\ref{ineq-constraint-1}) is inactive and (\ref{ineq-constraint-2}) is
		active}\}. 
\end{align*}
Note that it is impossible that both inequality constraints are active at the same time.
It follows that
\[
S \cup S_1 \cup S_2 =\{1, 2, \ldots, m\}
\]
and the three sets are mutually exclusive.  From the definition in (\ref{def-F}),
$\nabla F(\bfx)$ is a $1\times m$ row vector whose $i$-th entry is
\beq{grad-F}
\left(\nabla F(\bfx)\right)_i = \frac{2(x_i-\eta_i)}{m-1}.
\eeq
From the definition in (\ref{def-h}), $\nabla h(\bfx)$ is a $1\times m$ row vector
whose entries are all one, \ie
\beq{grad-h} 
\left(\nabla h(\bfx)\right)_i=1.
\eeq 
From the definitions in (\ref{def-g1}) and 
(\ref{def-g2}), both $\nabla \bfg_1(\bfx)$ and $\nabla \bfg_2(\bfx)$ are $m\times m$
matrices.  It is easy to verify that 
\beq{grad-g}
\nabla \bfg_1(\bfx)=-\bfI, \quad \nabla\bfg_2(\bfx)=\bfI,
\eeq
where $\bfI$ denotes the $m\times m$ identity matrix.
From \req{grad-F}, \req{grad-h} and \req{grad-g}, one can simplify (\ref{KKT}) to 
obtain
\beq{KKT-1}
x_i^* =\eta_i -\frac{m-1}{2}(\lambda-\mu_i+\omega_i),
\eeq
where $\mu_i$ and $\omega_i$ are the $i$-th entry of $\bfmu$ and $\bfomega$, respectively.

Now we derive $\lambda$, $\bfmu$ and $\bfomega$.
One can rewrite constraint (\ref{eq-constraint})
as follows.
\begin{align}
	L = \sum_{j=1}^m x_j^* 
	&= \sum_{j\in S} x_j^* +\sum_{j\in S_1}x_j^* +\sum_{j\in S_2}x_j^* \nonumber\\
	&= \sum_{j\in S} x_j^* + 0 + \sum_{j\in S_2} r_j \nonumber \\
	&= \sum_{j\in S} \eta_j - \frac{m-1}{2}\lambda\cdot |S|
	+ \sum_{j\in S_2} r_j,\label{intermed-1}
\end{align}
where the last equality is due to \req{KKT-1} and $|S|$ denotes the number of elements
in set $S$. From (\ref{intermed-1}), we have
\beq{lambda}
\lambda=\frac{\sum_{j\in S}\eta_j + \sum_{j\in S_2} r_j-L}{(m-1)|S|/2}.
\eeq
Now we determine $\bfmu$ and $\bfomega$.  For index $i$, if
\[
0< \eta_i-(m-1)\lambda/2 < r_i,
\]
both inequality constraints are satisfied and are inactive.  In this case,
$\mu_i=0$ and $\omega_i=0$.
If $\eta_i-(m-1)\lambda/2\le 0$, constraint (\ref{ineq-constraint-1}) is violated and
constraint (\ref{ineq-constraint-2}) is satisfied.  In this case,
\begin{align}
\omega_i &=0 \nonumber\\
\mu_i &=-\frac{2\eta_i}{m-1}+\lambda.\label{mu_i}
\end{align}
Clearly, $\mu_i > 0$.  Finally, if
$\eta_i-(m-1)\lambda/2 \ge r_i$,
constraint (\ref{ineq-constraint-1}) is satisfied and
constraint (\ref{ineq-constraint-2}) is violated.  In this case,
\begin{align}
\mu_i &=0\nonumber\\
\omega_i &=\frac{2(\eta_i -r_i)}{m-1}-\lambda.\label{omega_i}
\end{align}
Clearly, $\omega_i > 0$.

In the following proposition, we present some properties of the optimal point given in
\req{KKT-1}.  The proof of the proposition is presented in Appendix C.  This proposition
suggests a method to find the optimal solution in \req{KKT-1}.  This method is called
water-filling algorithm and will be presented in the next section.
\bprop{opt-prop}
Let $\bfx^*$ be the optimal solution of \req{vcnp}.  
\begin{enumerate}
	\item Suppose that $z_i+r_i \ge z_j+r_j$, then 
\beq{imp1}
x_i^*=r_i \ \mbox{implies that}\ x_j^*=r_j.
\eeq
\item Suppose that $z_i \ge z_j$, then
\beq{imp2} 
x_j^*=0 \ \mbox{implies that}\ x_i^*=0.
\eeq
\item Suppose that $0<x_i^* < r_i$ and $0<x_j^*<r_j$, then
\beq{imp3}
x_i^* +z_i=x_j^* +z_j.
\eeq
\end{enumerate}
\eprop

\bsubsec{A water-filling algorithm}{WFA}

\begin{figure}[htb]
	\centerline{\includegraphics[width=\linewidth]{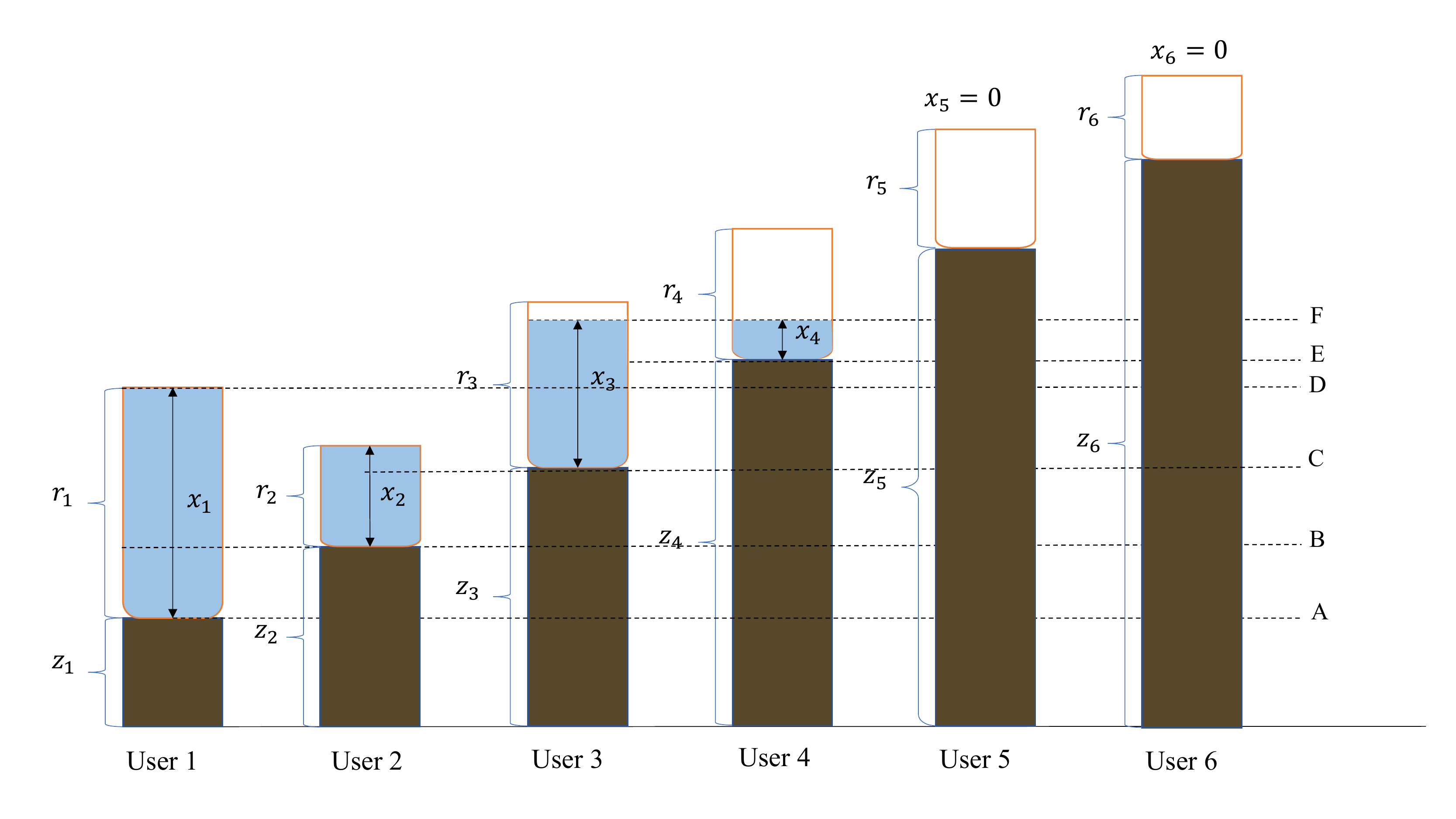}}
	\caption{Illustration of a water-filling problem.  There are six 
		buckets sitting on six bins of heights $z_1, z_2, \ldots, z_6$.  
		The bucket sizes are $r_1, r_2, \ldots, r_6$.  
		Fill $L$ units of water into the buckets starting from the 
		buckets whose bottoms are the
		lowest until the water is exhausted.  That is, $x_1+x_2+\ldots+x_6=L$. 
		In this example, buckets 1 and 2 are completely full.  Buckets 3 and 
		4 are partially filled. Bucket 5 and bucket 6 are completely empty.}
	\label{wf}
\end{figure}

In this section we propose an algorithm to find the optimal point $\bfx^*$ and
sets $S, S_1$ and $S_2$.  \rprop{opt-prop} at the end of the last section provides
a view to the optimization problem \req{vcnp}.  
One can view bits as water, requests as buckets, the number of 
cumulatively granted bits as the height
of bins on the top of which buckets sit, and  the
total number of bits that can be transmitted as the amount of water.
In this view, the solution of problem \req{vcnp} is a method of filling water into buckets
of finite sizes, sitting on tops of bins of various heights.
Specifically, let there be $m$ buckets.  
The sizes of the buckets
are $r_1, r_2, \ldots, r_m$.  The $j$-th bucket sits on the top of a bin of height $z_j$.  
There are $L$ units of water.  The goal is to fill water into the buckets such that
the height of water surfaces in buckets are as even as possible.
See Fig. \ref{wf} for a graphical illustration.
Statements (1) and (2) of \rprop{opt-prop} imply that water starts to fill
buckets on lower bins first, and buckets with higher tops are full only after
buckets with lower tops are full.  Statement (3) says that partially filled
buckets have the same water level.  We refer the reader to Fig. \ref{wf} 
for an example.  Finding the optimal solution of \req{vcnp} is equivalent to
filling water into buckets simultaneously starting from buckets on lowest bins.
Once a bucket is full, it stops receiving water, and water is filled into other buckets
until it is exhausted. This water-filling problem can easily be solved by a recursive
algorithm shown in Algorithm \ref{wfa}.   This algorithm works by raising the water level
from one possible level to the next higher possible level.  There are three possible water
levels in the algorithm.  The possible water levels are 
\begin{enumerate}
	\item bottoms of one of the buckets;
	\item tops of one of the buckets; 
	\item water level at a partially filled bucket.
\end{enumerate}

Let $P(\bfz, \bfr, m, L)$ denote the recursive function in Algorithm \ref{wfa}.
The output of this function is the optimal point of the optimization problem
\req{vcnp}, \ie
\[
\bfx^*=P(\bfz, \bfr, m, L).
\]
Let 
\[
j=\argmin_i\{z_i : r_i > 0, 1\le i\le m\}.
\]
User $j$ is said to be least granted.

Let $\N$ be the set of users whose requests are non-zero, \ie
\[
\N=\{i : r_i > 0, 1\le i\le m\}.
\]
Then, sort $z_i, i \in \N$, into an ascending sequence.  Label user's identities such
that 
\beq{ordering-1}
z_1=z_2=\ldots=z_k < z_{k+1}\le \ldots\le z_{|\N|}
\eeq
for some $k$, where $1\le k\le |\N|$, and such that
\beq{ordering-2}
r_1\le r_2\le \ldots\le r_k.
\eeq
There are $k$ eMBB users that are least granted
in the number of bits.  Let
\beq{minimum}
\tilde m=\min\{r_1, z_{k+1}-z_k\}.
\eeq
If $L\le \tilde m k$, we will show in the proof of \rthe{optimal-wf} to be presented
below that the optimal solution is
\[
x_j=\left\{\begin{array}{ll}
	L/k & 1\le j\le k \\
	0 & k+1\le j\le |\N|.\end{array}\right.
\]
If $L> \tilde m k$, we raise the water level by $\tilde m$.
We introduce a new set of parameters $\bfz'$, $\bfr'$, $L'$, and compute
\beq{new-prob}
\bfy=P(\bfz', \bfr', m, L'),
\eeq
where
\begin{align}
	z_\ell' &= \left\{\begin{array}{ll} z_\ell +\tilde m & 1\le \ell\le k \\
		z_\ell & k+1\le \ell \le |\N| \end{array}\right.  \label{z'}\\
	r_\ell' &= \left\{\begin{array}{ll} r_\ell -\tilde m & 1\le \ell\le k \\
		r_\ell & k+1\le \ell \le |\N| \end{array}\right.  \label{r'}\\
	L' &= L-\tilde m k.\label{L'}
\end{align}
The optimal point $\bfx$ of the original problem is related to $\bfy$.  There are two cases
depending on how the minimum in \req{minimum} is achieved.
In the first case, $r_1< z_{k+1}-z_k$.  We assume that there are $j$ users whose
requests are the same as $r_1$, \ie
\[
\tilde m=r_{1}=r_{2}=\ldots=r_{j}
\] 
for some $1\le j\le k$. In this case,
the optimal point of the original problem $\bfx$ is related to $\bfy$ through
\beq{opt-x-1}
x_\ell=\left\{\begin{array}{ll}
	\tilde m & \ell=1, 2, \ldots, j \\
	y_\ell +\tilde m & \ell=j+1, j+2, \ldots, k \\
	y_\ell & k+1\le \ell \le |\N|. \end{array}\right.
\eeq
In the second case, $r_1\ge z_{k+1}-z_k$ and
\[
\tilde m=z_{k+1}-z_k.
\] 
In this case, the optimal point of the original problem $\bfx$ is related to $\bfy$ through
\beq{opt-x-2}
x_\ell=\left\{\begin{array}{ll}
	y_\ell + \tilde m & 1\le \ell \le k \\
	y_\ell & k+1\le \ell \le |\N|. \end{array}\right.
\eeq

The following theorem states that the water-filling algorithm solves the constrained 
optimization problem in \req{cnp}.  Its proof is presented in Appendix C.
\bthe{optimal-wf}
If $L\le \sum_{j=1}^m r_j$, the water-filling algorithm finds the optimal 
point $\bfx^*$ that satisfies \req{KKT-1}, \req{lambda}, (\ref{mu_i}) and (\ref{omega_i}).
\ethe

We remark that the idea of viewing resources as water and allocating resources
as if filling water in buckets have been used in several resource allocation problems
\cite{Radunovic2007,Coluccia2012}.  In fact, several algorithms that solve resource
allocation problems are also called water-filling algorithms.  There are two types of
water-filling problems in the literature.  We refer the reader to Fig. 1 in \cite{Coluccia2012}.
The first type, shown in Fig. 1(a) of the reference, arises
in a power control problem of Gaussian interference channels \cite{Shum2007, Yu2002}.
However, the corresponding optimization problem has a logarithmic objective function.
In addition, the assigned power has no upper limit.  
The corresponding water-filling problem can be viewed as a special case of our problem, 
in which the buckets have infinite capacities.   The water-filling problem in Fig. 1(b)
of \cite{Coluccia2012} arises in a max-min fairness allocation problem.  The
water-filling problem is also a special case of ours, in which the heights of bins 
$\{z_j: 1\le j\le m\}$ are all zero.  We refer the reader to \cite{Radunovic2007} for more
information on max-min fairness allocation.

\begin{algorithm}
	\caption{Water filling algorithm\label{wfa}}
	\raggedright {\bf Function:} $\bfx=P(\bfz, \bfr, m, L)$\newline
	\raggedright {\bf Inputs}: $\bfz, \bfr, m$, and $L$\newline
	\raggedright {\bf Outputs}: $\bfx$
	
	\begin{algorithmic}[1]
	\State Let $\N=\{i : r_i >0\}$.
	\State Label the users in set $\N$ and let $k$ be such that
	\[
		z_1=z_2=\ldots=z_k < z_{k+1}\le \ldots\le z_{|\N|}.\]	
	\State In addition, label users $1, 2, \ldots, k$ such that
	\[
	r_1\le r_2\le \ldots\le r_k.
	\]
\State Let $\tilde m=\min\{r_1, z_{k+1}-z_k\}$.
\If{$L\le \tilde m k$}
\State $x_1=x_2=\ldots=x_k=L/k$
\State $x_j=0, \ j=k+1, k+2,\ldots, |\N|$
\State \Return $\bfx$
\EndIf
\State	Evaluate $\bfz'$ according to (\ref{z'}). 
\State	Evaluate $\bfr'$ according to (\ref{r'}).
\State	Evaluate $L'$ according to (\ref{L'}). 
%\State  Update $\N$.
\State Call $\bfy=P(\bfz', \bfr', m, L')$.
\If{$r_{1} < z_{k+1}-z_k$}
\State Compute $\bfx$ according to \req{opt-x-1}.
\ElsIf{$z_{k+1}-z_k \le r_1$}
\State Compute $\bfx$ according to \req{opt-x-2}.
\EndIf
\State \Return $\bfx$
%\EndWhile
\end{algorithmic}
\end{algorithm} 

\bsec{Numerical and simulation results}{numerical}

In this section we present numerical and simulation results.  We present the
results in the following two sub-sections.  
In \rsubsec{num-pei} we examine the light traffic
analysis of $\pr(E_i)$ by comparing numerical results with simulation results. 
In \rsubsec{res-alloc}, we study the Nash equilibria of the two-player
resource allocation games and the performance of the water-filling algorithm. 

\bsubsec{Light traffic approximation}{num-pei}

In this section we first examine the light traffic
analysis of $\pr(E_i)$ by comparing numerical results with simulation results.
Then, we study the effect of $p$ and $\tau$ to the packet loss probability
$\pr(E_i)$.  

\begin{figure}[htb]
	%\centering
	\centerline{\includegraphics[width=\linewidth]{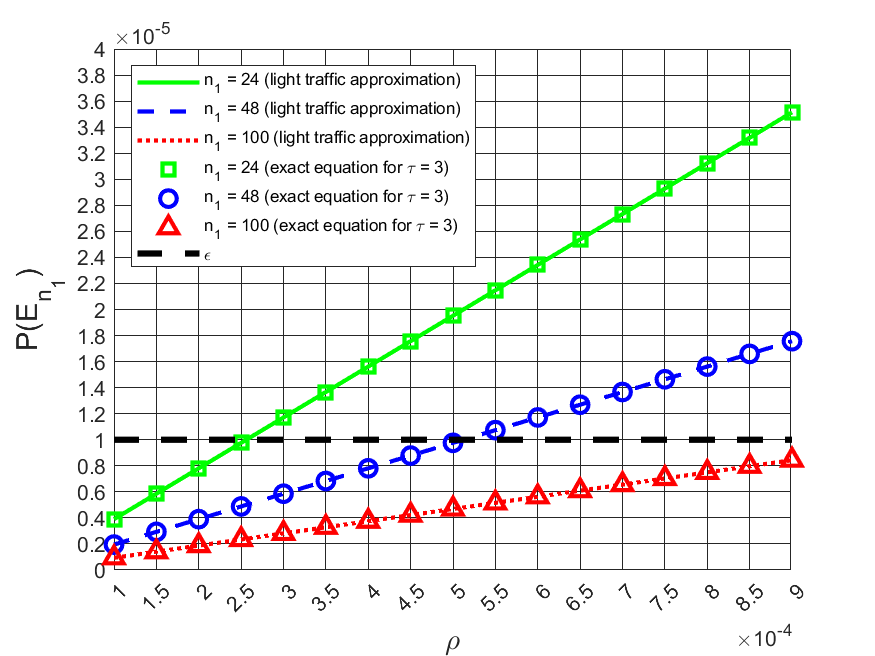}}
	\caption{$\pr(E_{n_1})$ versus $\rho$.}\label{various-rho}
\end{figure}

\begin{figure}[htb]
	%\centering
	\centerline{\includegraphics[width=\linewidth]{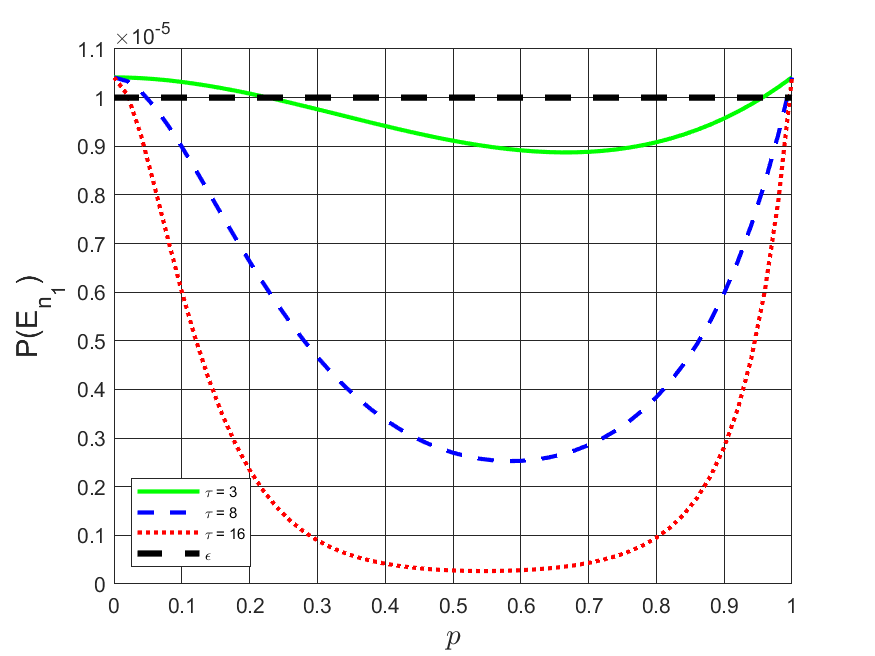}}
	\caption{$\pr(E_{n_1})$ versus $p$.}\label{various-p}
\end{figure}

In Fig. \ref{various-rho} we compare the light traffic approximation with 
$\tau=3$ in (\ref{pek}) and its exact result in (\ref{final-tau=3}). We assume that
the retransmission probability $p$ for URLLC packets is 0.3.  We consider three
sizes for the common region: $i=24$ for a small common region, $i=48$ for a medium
sized common region and $i=100$ for a large common region.  From Fig. \ref{various-rho}
we see that the light traffic approximation and the exact result are very close to 
each other.  Next we show $\pr(E_i)$ versus $p$ in Fig. \ref{various-p}.  
As expected, $\pr(E_i)$ decreases with $p$ initially.  Later it increases with $p$.
There is an optimal value of $p$ at which $\pr(E_i)$ achieves a minimal value.

\bsubsec{Resource allocation}{res-alloc}

\begin{figure}[htb]
	%\centering
	\centerline{\includegraphics[width=\linewidth]{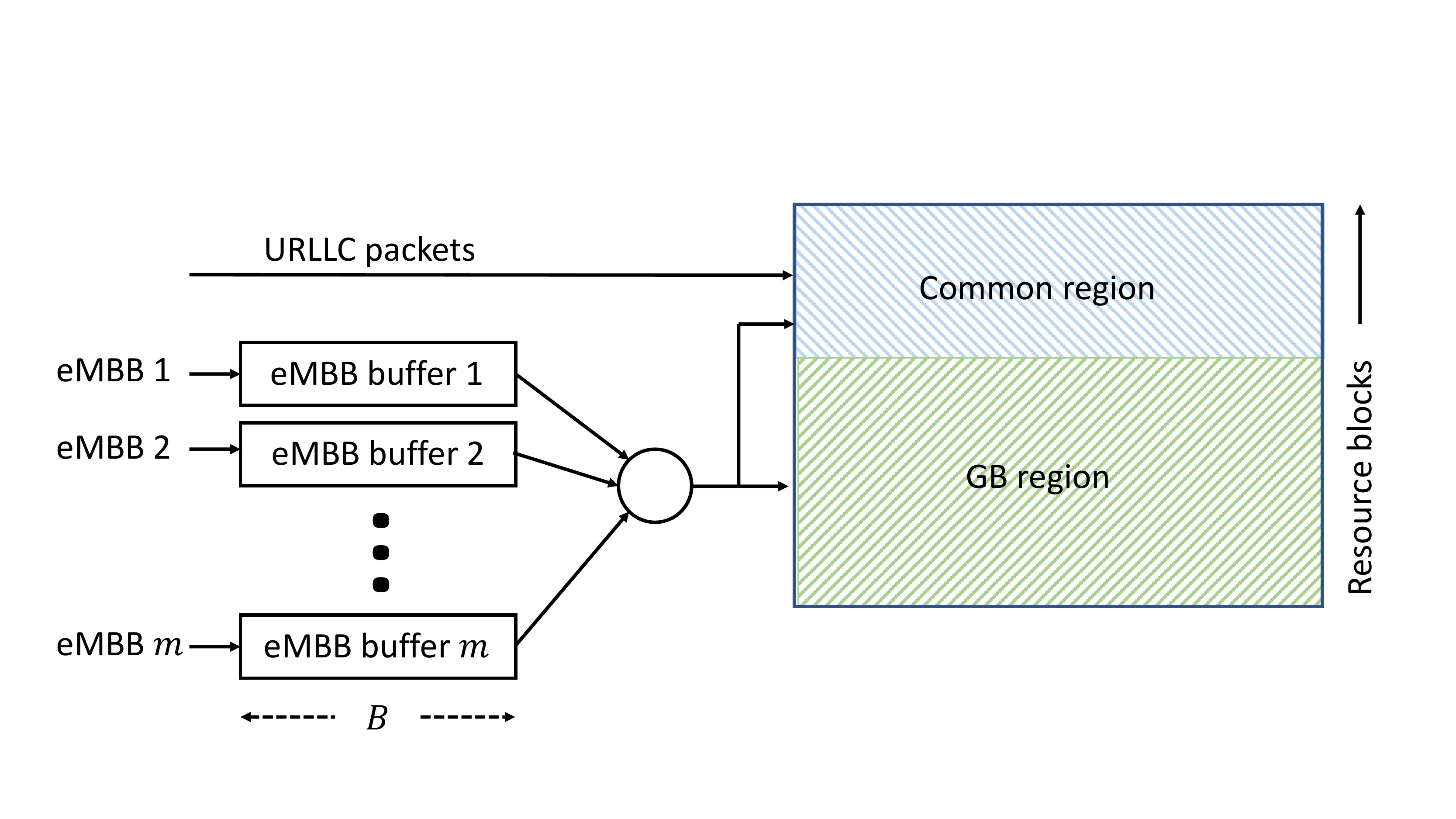}}
	\caption{System architecture.}\label{simu-arch}
\end{figure}

In this section we study the two resource allocation algorithms  
presented in \rsec{two-player-game} and \rsec{np}, respectively.   
We simulate the system shown in Fig. \ref{simu-arch}.  The sizes of common region
and grant-based region are determined by a resource allocation algorithm.
Each eMBB user has a buffer which can store $B$ eMBB bits.  eMBB packet arrivals
in frame $t$ are stored in the corresponding buffer.  If the residual space of a
buffer is not enough to store new arrivals, eMBB packets can be lost.
At the beginning of frame $t+1$,
eMBB users make a request to transmit all bits in their buffers.  A resource allocation
algorithm determines the number of bits that each user can transmit.  Packets
that can not be transmitted stay in the buffers.  
We will compare the water-filling algorithm with a few other resource allocation algorithms.  

\begin{figure}[htb]
	%\centering
	\centerline{\includegraphics[width=\linewidth]{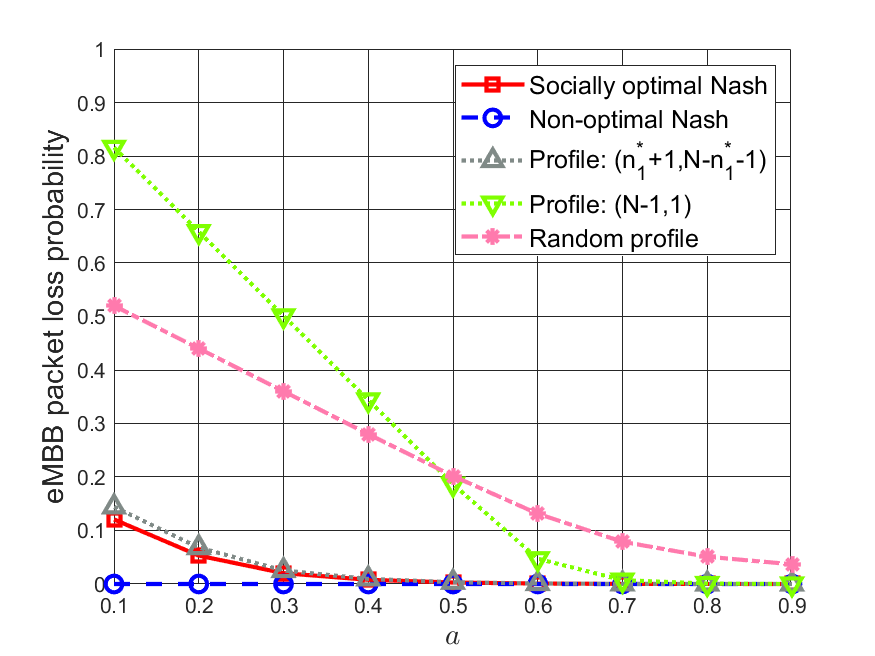}}
	\caption{Packet loss probability of eMBB traffic versus $a$.}\label{plpe}
\end{figure}

\begin{figure}[htb]
	%\centering
	\centerline{\includegraphics[width=\linewidth]{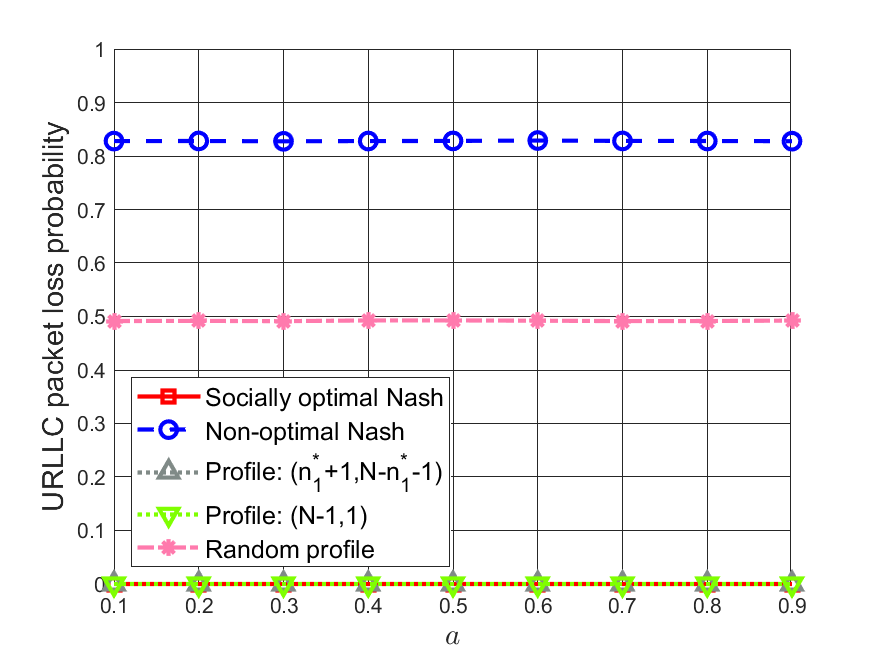}}
	\caption{Packet loss probability of URLLC traffic versus $a$.}\label{plpu}
\end{figure}

\begin{figure}[htb]
	%\centering
	\centerline{\includegraphics[width=\linewidth]{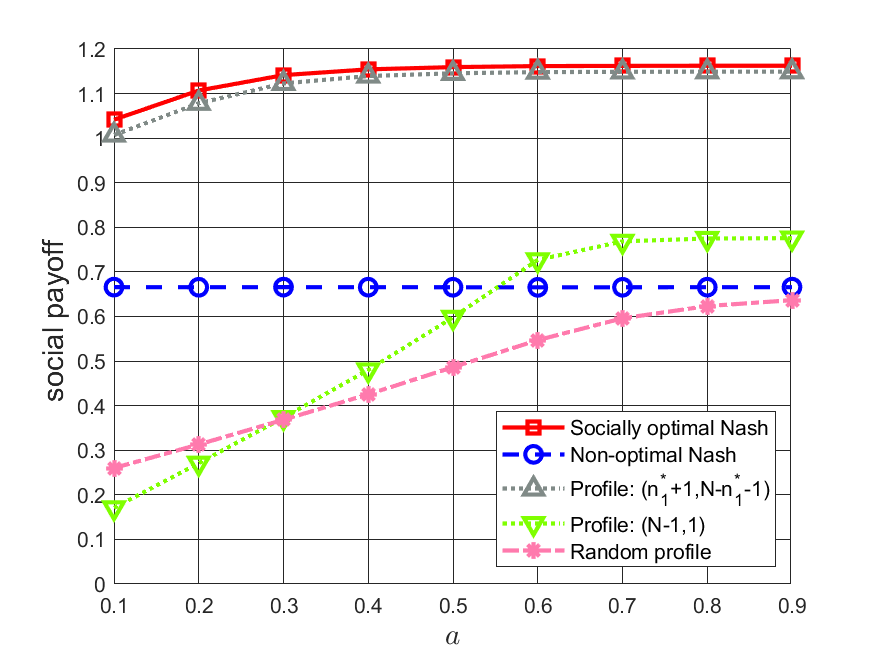}}
	\caption{Social payoffs versus $a$.}\label{sumpayoff}
\end{figure}

First, we study the packet loss probability of eMBB traffic 
when wireless resources are assigned according to the socially optimal Nash equilibrium.
We assume that eMBB bits are scheduled using the water-filling algorithm
in Algorithm \ref{wfa}.  We assume that the number of bits to 
each eMBB buffer in a time frame is
an independent discrete uniform random variable over $[0, 3\times10^5]$.  
Values of other parameters are shown in Table \ref{Table-2}.
We compare the performance with that of a few other resource allocation strategies.
Specifically, we compare the performance of socially optimal Nash equilibrium 
$(n_1^*, n_2^*(n_1^*))$ with
the non-optimal Nash equilibrium $(0, n_2^*(0))$, profiles $(n_1^*+1, N-n_1^* -1)$ and
$(N-1, 1)$, and a random profile.  In time slot $t$, the random profile assigns $N_{1,t}$
resource blocks to the common region, and assigns $N_{2,t}$ resource blocks to the grant-based
region.  Discrete random variable $N_{1,t}$ is uniform over the interval $[0, N]$.
Conditioning on $N_{1,t}$, discrete random variable $N_{2,t}$ is uniform over the interval
$[0, N-N_{1,t}]$.  The packet loss probability of eMBB traffic
is shown in Fig. \ref{plpe}.  The packet loss probability of URLLC traffic is shown
in Fig. \ref{plpu}.  From these two figures, we see that the resource allocation based
on socially optimal profile $(n_1^*, n_2^*(n_1^*))$ achieves a good balance between
the two loss probabilities.  We have also simulated the social payoff, which is defined
as the sum of payoffs of two game players.  The result is shown in Fig. \ref{sumpayoff}.
As expected, the socially optimal Nash equilibrium has the largest social payoffs.

\begin{table}[!htb]
	\begin{center}
	\begin{tabular}{|l|r|l|r|}\hline
		\textbf{Parameter} & \textbf{Value} & \textbf{Parameter} & \textbf{Value} \\\hline
		$\rho$ & $6.5\times 10^{-4}$ & $\tau$ & 8 \\ 
                  &packets per & & \\ 
		 &mini-slot & & \\ \hline
		$p$ & 0.3 & $N$ & 60 \\ \hline
		$\epsilon$ & $10^{-5}$ & $n_1^*$ & 30 \\ \hline
		$b$ & 0.8 & $c$ & $3.2\times10^4$ bits\\ \hline
		$B$ & $3.8\times10^5$ bits & $m$ & 8 \\ \hline
	\end{tabular}
\end{center}
\medskip
\caption{Parameter values.}\label{Table-2}
\end{table}

Finally, we compare the water-filling algorithm with a few other
algorithms.  In this study, we assume that resource blocks are partitioned
into common region and grant-based region according to the profile $(n_1^*, n_2^*(n_1^*))$.
Besides the water-filling algorithm, other resource allocation algorithms that we consider
are smallest request first, largest request first, random order allocation, two-step 
averages, and max-min fairness allocation.  We now briefly explain these allocation methods.
The smallest request first and the largest request first methods are 
straightforward.  Sort the requests $\{r_j: 1\le j\le m\}$ into an ascending list.  
In the case of smallest request first, grant requests in
the ascending order.  On the other hand, in the case of largest request first, 
grant requests in the descending order.  In either case, grant requests 
until the allowance $L$ is exhausted or all requests
have been satisfied. The random order allocation is similar.  Arrange requests according to
a randomly selected permutation and grant requests according to the order of this permutation
until $L$ bits are exhausted or all requests are satisfied.  
The two-step average method works as follows.  There are two steps in this 
method. In the fist step, compute and grant requests according to the average 
$v=L/m$.  The number of bits granted to user $i$ in step 1 is
$x_i^{(1)}=\min(r_i, v)$.  The residual request size of user $i$ is 
$r_i^{(1)}=r_i-x_i^{(1)}$.  The original grant allowance is $L$
and the residual allowance after the first step is $L^{(1)}=L-\sum_{i=1}^m x_i^{(1)}$.
In the second step, sort $\{r_i^{(1)}: 1\le i\le m\}$ into a descending list.  Grant requests
in the order of this list until $L^{(1)}$ is exhausted or all requests have been satisfied.
Finally, we explain the max-min fairness method.  This allocation method was proposed to allocate
bandwidth \cite{Radunovic2007}.  The max-min fairness allocation can be solved by a 
water-filling problem.  This problem is a special case of $P(\bfz, \bfr, m, L)$, in which
$\bfz$ is set to a zero vector, \ie $\bfz=\bfzero$. Although the max-min allocation can be solved
by a water-filling algorithm, there is a simpler algorithm to solve it.  First, sort
$\{r_i: 1\le i\le m\}$ into an ascending sequence.  Relabeling indexes 
if necessary, we can assume that the sequence $\{r_i: 1\le i\le m\}$ is ascending.  
For convenience, let $r_i^{(0)}=r_i$ for $i=1, 2, \ldots, m$, and let 
$L^{(0)}=L$.  In step $j$, compute the average number of bits that can be granted
to $j$ users, \ie
\[
v^{(j)}=L^{(j-1)}/(m-j+1).
\]
In step $j$, allocate 
\iffalse
\[
x_i^{(j)}=\left\{\begin{array}{ll} \min(v^{(j)},r_i^{(j-1)}) & \mbox{if $j\le i\le m$}\\
	0 & \mbox{if $1\le i\le j-1$}\end{array}\right.
\] \fi
\[
x_i^{(j)}=\min(v^{(j)},r_i^{(j-1)})
\]
to user $i$.  Calculate the residual request sizes for the next step, \ie 
\[
r_i^{(j)}=r_i^{(j-1)}-x_i^{(j-1)},\qquad 1\le i\le m.
\]
If $r_i^{(j)}=0$ for all $i$, the algorithm stops.  Otherwise, the algorithm continues and 
we compute the residual allowance 
\[
L^{(j)}=L^{(j-1)}-\sum_{i=1}^m x_i^{(j)}.
\]

\begin{figure}[htb]
	%\centering
	\centerline{\includegraphics[width=\linewidth]{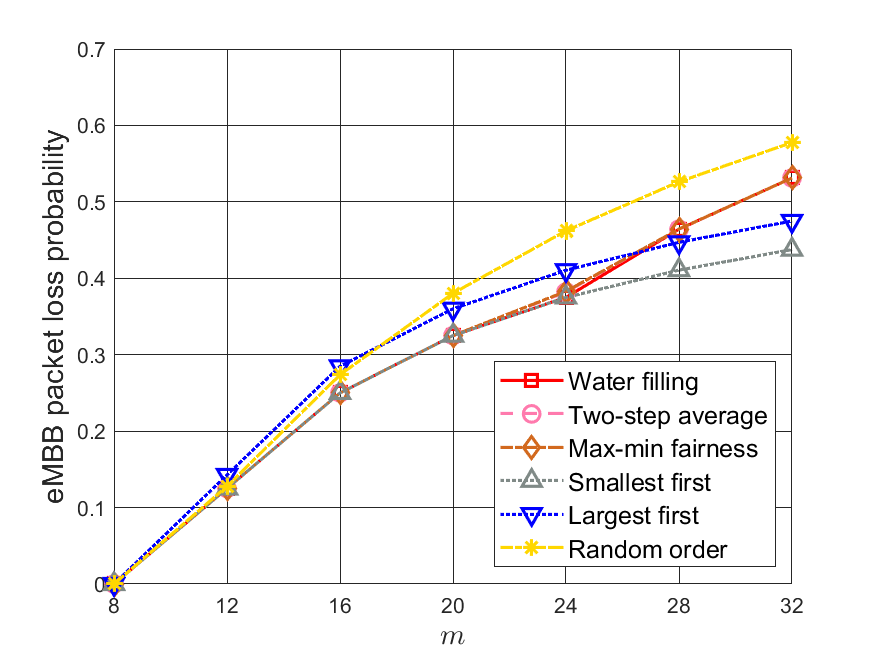}}
	\caption{eMBB packet loss probability versus $m$.}\label{palo2}
\end{figure}

\begin{figure}[htb]
	%\centering
	\centerline{\includegraphics[width=\linewidth]{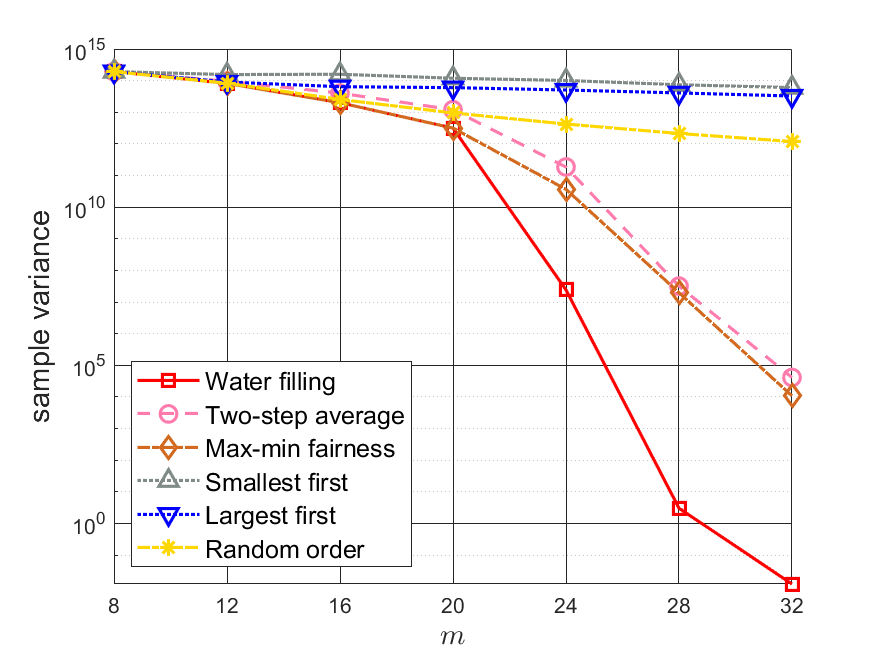}}
	\caption{Sample variance versus $m$.}\label{samv2}
\end{figure}

\begin{figure}[htb]
	%\centering
	\centerline{\includegraphics[width=\linewidth]{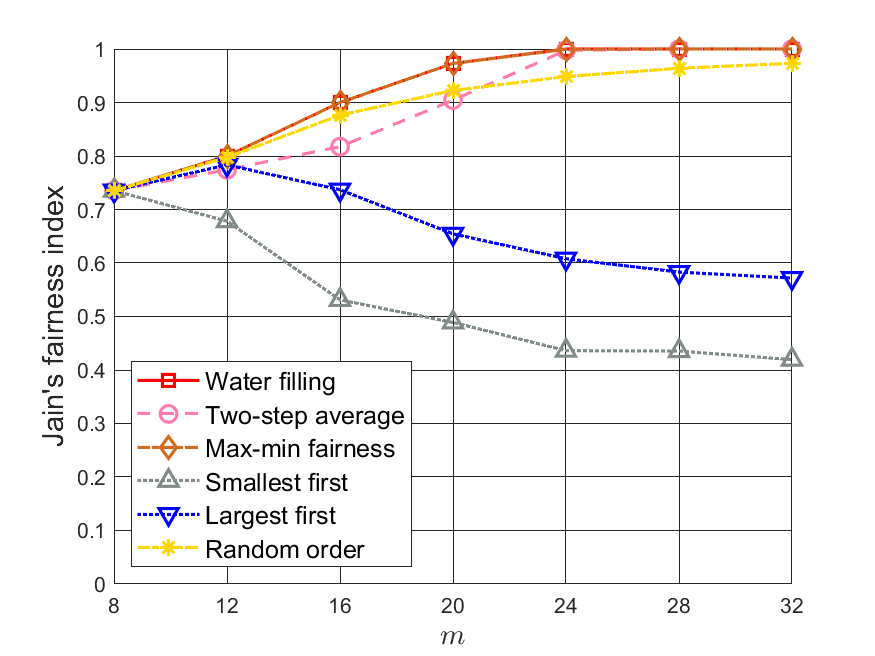}}
	\caption{Jain's fairness index versus $m$.}\label{jain2}
\end{figure}

We simulate the water-filling algorithm and the five other resource allocation algorithms
for $T=10^6$ frames.  The sample variance of a sequence $\{z_i(t+1): 1\le i\le m\}$ is defined
in \req{var}.  Another well known measure of fairness is the Jain's fairness index defined
as
\[
\frac{\left(\sum_{i=1}^m z_i(t+1)\right)^2}{m\cdot \sum_{i=1}^m z_i(t+1)^2}.
\]
The eMBB packet loss probability, sample variance (defined in \req{var})
and Jain's fairness index in Figs. \ref{palo2}, \ref{samv2} and \ref{jain2}, respectively.
From Fig. \ref{palo2}, the smallest request first method has the smallest eMBB packet loss
probability.  The water-filling algorithm and the max-min fairness method have very 
similar results.  From Figs. \ref{samv2} and \ref{jain2}, the water-filling algorithm has
the best performance in both sample variance and the Jain's fairness index.


\bsec{Conclusions}{conclusions}

In this paper we studied a wireless resource allocation problem and a packet scheduling problem
of the URLLC traffic and eMBB traffic in an uplink wireless network.
We proposed to divide frequencies into a grant-based region for eMBB traffic 
only, and a common region for both URLLC traffic and eMBB traffic.  
To cope with the ultra stringent latency and reliability requirement of URLLC packets,
we proposed a persistent random retransmission scheme for URLLC packets.
We derived the probability that a randomly selected URLLC packet fails to meet its
latency requirement.  We proposed a two-player game to determine the sizes of the common
region and the grant-based region.  We studied the Nash equilibria of this game.
For the request-and-grant allocation of eMBB packets, we proposed a constraint
nonlinear program which minimizes the variance of the number of bits 
granted to the eMBB users.  We show that a water-filling algorithm can recursively solve 
the nonlinear program. From simulation, we show that our scheme, consisting of
resource allocation according to Nash equilibria of a game, persistent random transmission
of URLLC packets and scheduling eMBB packets by a water-filling algorithm, works 
better than four other heuristic methods.

\bigskip

\begin{center}
	{\bf Appendix A}
\end{center}
{\bf Proof of \rprop{incr}.}
Let $X_j$ denote the number of Poisson 
arrivals in mini time slot $j$, $-(\tau-1)\le j\le \tau-1$, not including the
tagged arrival in mini time slot $0$.  We use bold face letters to denote
vectors.  Let $\bfx$ be a $1\times (2\tau-1)$ row vector, \ie
\[
\bfx=(x_{-(\tau-1)}, \ldots, x_{-1}, x_0, x_1, \ldots, x_{\tau-1}).
\]
Define function $f_i(\bfx)$ according to (\ref{G_0}) and (\ref{G_i})
\begin{align}
	&f_i(\bfx) = \nonumber\\
	&\left\{\begin{array}{ll}
1-\indicatorFunction{x_0=0}(1-p)^{\sum_{i=-(\tau-1)}^{-1} x_i} & \mbox{if $i=0$}\\
1-p\cdot \indicatorFunction{x_i=0}(1-p)^{\sum_{j=i-(\tau-1)}^{i-1} 
	x_j} & \mbox{if $1\le i\le \tau-1$.}\end{array}\right.		
\end{align}
Also define function
\[
f(\bfx)=\prod_{i=0}^{\tau-1} f_i(\bfx).
\]
It is easy to verify that $f_i(\bfx)$ is non-decreasing with respect to 
$x_j$ for any $j=-(\tau-1), \ldots, \tau-1$.   Thus, $f(\bfx)$ is also non-decreasing.
Let $\bfX=(X_{-(\tau-1)}, \ldots, X_{-1}, X_0, X_1, \ldots, X_{\tau-1})$, where
$X_i$ is the number of Poisson arrivals in mini time slot $i$.  One can
rewrite (\ref{total-prob}) as follows,
\begin{align}
\ex[\indicatorFunction{E}\;|\; \bfX] &= \prod_{i=0}^{\tau-1} f_i(\bfX)
=f(\bfX)\nonumber\\
\pr(E) &= \ex[\ex[\indicatorFunction{E}\;|\; X]]=\ex[f(\bfX)].\nonumber
\end{align}
Let $\bfY=\{Y_j: -(\tau-1)\le j\le \tau-1\}$ be a sequence of 
independent Poisson random variables with mean $\rho'$, where $\rho' \ge \rho$.
It is well known that Poisson random variables are stochastically increasing
in means (see Example 9.2(B) on page 411 of Ross \cite{RosS.96}).
Thus,
\[
Y_j \stge X_j
\]
for all $j=-(\tau-1), \ldots, \tau-1$.  By Example 9.2(A) of 
Ross \cite[p. 410]{RosS.96}, 
\[
\ex[f(\bfY)] \ge \ex[f(\bfX)].
\]
It follows that $\pr(E)$ increases with $\tilde\rho$.

\begin{center}
	{\bf Appendix B}
\end{center}
{\bf Proof of \rthe{Nash}.}  Suppose that $(n_1, n_2)$ is an action profile of the
two-player resource allocation game. Recall that profile $(n_1, n_2)$ is 
called a pure Nash equilibrium
if action $n_1$ and action $n_2$ are best responses to each other.  
\iffalse
Suppose that $\min_{0\le i\le N} \pr(E_i)> \epsilon$.  We now
show that $(0, n_2^*(0))$ is the only one pure Nash equilibrium.  To prove this, we show that
any action profile $(i, j)$, where $(i,j)\ne (0, n_2^*(0))$, is not a Nash equilibrium.
That is, either $i$ is not a best response to $j$, or $j$ is not a best response to 
$i$.  For any profile $(i, j)$, such that $i > 0$ and $0\le j\le N$, action zero
of the URLLC player is a better response against action $j$, because
\begin{align*}
&P_{URLLC}(0,j)=\left(0-\frac{0}{N}\cdot b\right)+1=1 \\
&\quad >\left(0-\frac{i}{N}\cdot b\right)+1=P_{URLLC}(i,j).
\end{align*}
Then consider profile of the form $(0, j)$.  We show that $n_2^*(0)$ is
a better response than any other action $j$ of the eMBB player. Indeed,
for $j<n_2^*(0)$, we have
\begin{align*}
&P_{eMBB}(0,n_2^*(0))=\left(\frac{n_2^*(0)}{N}\cdot 1-\frac{n_2^*(0)}{N}\cdot b\right)+1 \\
&\quad >\left(\frac{j}{N}\cdot 1-\frac{j}{N}\cdot b\right)+1=P_{eMBB}(0,j).
\end{align*}
For $j>n_2^*(0)$, we have
\begin{align*}
&P_{eMBB}(0,n_2^*(0))=\left(\frac{n_2^*(0)}{N}\cdot 1-\frac{n_2^*(0)}{N}\cdot b\right)+1 \\
&\quad>\left(\frac{n_2^*(0)}{N}\cdot 1-\frac{j}{N}\cdot b\right)+1=P_{eMBB}(0,j).
\end{align*}
\fi

Now we consider the case, in which $\min_{0\le i\le N} \pr(E_i)\le \epsilon$.
There are three sub-cases.  Since their proofs are similar, we only present the proof
of the second sub-case.  Proofs of the other sub-cases are omitted.  We also omit
the proof of the case, in which $\min_{0\le i\le N} \pr(E_i)> \epsilon$.
Suppose that \req{case2} holds.  We shall show that profiles
$(n_1^*, n_2^*(n_1^*))$ and $(0, n_2^*(0))$ are pure Nash equilibria.
We also need to show that profile $(i, j)$ is not a Nash equilibrium if
$(i, j)\ne (n_1^*, n_2^*(n_1^*))$ and $(i, j)\ne (0, n_2^*(0))$.  
In addition, we shall discuss the social optimality of these two equilibria.

To begin, we show that profile $(n_1^*, n_2^*(n_1^*))$ is a pure Nash equilibrium. We show
that action $n_1^*$ of the URLLC agent is a best response to action $n_2^*(n_1^*)$
of the eMBB agent.  For $i<n_1^*$,
\beq{opt-payoff1}
\pu(n_1^*, n_2^*(n_1^*))=1-\frac{n_1^*}{N}b
\eeq
and
\[
\pu(i, n_2^*(n_1^*))=-\frac{i}{N}b.
\]
Since $0<b<1$, it follows that 
\beq{br1}
\pu(n_1^*, n_2^*(n_1^*))> \pu(i, n_2^*(n_1^*))
\eeq
for all $i < n_1^*$.  Now consider $i$, such that $i > n_1^*$.
We have
\begin{align}
\pu(i, n_2^*(n_1^*))&=1\cdot \indicatorFunction{i+n_2^*(n_1^*)\le N}-\frac{ib}{N}\nonumber\\
&=\left\{\begin{array}{ll} -\frac{ib}{N} & \mbox{if $i> N-n_2^*(n_1^*)$}\\
	1-\frac{ib}{N} & \mbox{if $i \le N-n_2^*(n_1^*)$}\end{array}\right.\label{subopt-payoff1}
\end{align}
Comparing \req{opt-payoff1} with (\ref{subopt-payoff1}), we obtain
\beq{br2}
\pu(n_1^*, n_2^*(n_1^*))> \pu(i, n_2^*(n_1^*))
\eeq
for all $i > n_1^*$.  \req{br1} and \req{br2} establish that action $n_1^*$ is a best response
to action $n_2^*(n_1^*)$.
We also need to show that action $n_2^*(n_1^*)$ is a best response to action $n_1^*$.  The payoff of
the eMBB agent when profile $(n_1^*, n_2^*(n_1^*))$ is taken is
\begin{align}
	\pe(n_1^*, n_2^*(n_1^*)) &=\frac{n_2^*(n_1^*)}{N}\indicatorFunction{n_1^*+n_2^*(n_1^*)\le N}-
	\frac{n_2^*(n_1^*)b}{N}\nonumber\\
	&= \frac{n_2^*(n_1^*)(1-b)}{N}.\label{opt-payoff2}
\end{align}
The payoff of the eMBB agent for an arbitrary action $j$ is
\[
\pe(n_1^*, j)=\left\{\begin{array}{ll} 
	\frac{n_2^*(n_1^*)}{N}\indicatorFunction{n_1^*+j\le N}-\frac{jb}{N} & \mbox{if $j>n_2^*(n_1^*)$}\\
	\frac{j}{N}\indicatorFunction{n_1^*+j\le N}-\frac{jb}{N} & \mbox{if $j\le n_2^*(n_1^*)$.}
	\end{array}\right.
\]
Since $n_1^*+n_2^*(n_1^*)\le N$, the preceding payoff is equal to
\beq{subopt-payoff2}
\pe(n_1^*, j)=\left\{\begin{array}{ll} 
	\frac{n_2^*(n_1^*)}{N}\indicatorFunction{n_1^*+j\le N}-\frac{jb}{N} & \mbox{if $j>n_2^*(n_1^*)$}\\
	\frac{j}{N}-\frac{jb}{N} & \mbox{if $j\le n_2^*(n_1^*)$.}
\end{array}\right.
\eeq
From (\ref{opt-payoff2}), \req{subopt-payoff2} and assumption $b<1$, we have
\[
 \pe(n_1^*, n_2^*(n_1^*)) \ge \pe(n_1^*, j).
\]
Hence, action $n_2^*(n_1^*)$ is a best response to action $n_1^*$.

Next, we show that profile $(0, n_2^*(0))$ is a pure Nash equilibrium.  First, we show
that action $0$ of the URLLC agent is a best response to action $n_2^*(0)$ of the eMBB agent.
Since 
\[
\pr(E_{\min(0, N-n_2^*(0))})=\pr(E_0)=1>\epsilon
\]
and $0+n_2^*(0)\le N$, it follows that
\beq{opt-payoff11}
\pu(0, n_2^*(0))=0.
\eeq
For arbitrary $i$,
\beq{temp1}
\pu(i, n_2^*(0))=\indicatorFunction{\pr(E_{\min(i,N-n_2^*(0))})\le \epsilon}
\indicatorFunction{i+n_2^*(0)\le N}-\frac{ib}{N}.
\eeq
We know that $n_1^*+n_2^*(0)>N$.
For $1\le i< n_1^*$, the first indicator function is $0$.  For $i \ge n_1^*$, the
second indicator function is $0$.  Thus, \req{temp1} is reduced to
\[
\pu(i, n_2^*(0))=-\frac{ib}{N}<\pu(0, n_2^*(0))
\]
for the both cases. Hence, action $0$ is a best response to action $n_2^*(0)$. 
We also need to show that action $n_2^*(0)$ is a best response to
action $0$.  Note that
\begin{align}
	\pe(0, n_2^*(0)) &= \frac{n_2^*(0)}{N}\indicatorFunction{0+n_2^*(0)\le N}-
	\frac{n_2^*(0)b}{N} \nonumber\\
	&=\frac{n_2^*(0)(1-b)}{N},\label{opt-payoff22}
\end{align}
since $n_2^*(0)\in [0, N]$.  On the other hand, the payoff of the eMBB agent for an arbitrary action $j$ is
\begin{align*}
\pe(0,j)&=\left\{\begin{array}{ll}
	\frac{n_2^*(0)}{N}\indicatorFunction{0+j\le N}-\frac{jb}{N} & \mbox{if $j>n_2^*(0)$}\\
	\frac{j}{N}\indicatorFunction{0+j\le N}-\frac{jb}{N} & \mbox{if $j\le n_2^*(0)$}
	\end{array}\right.\\
&=\left\{\begin{array}{ll}
	\frac{n_2^*(0)}{N}-\frac{jb}{N} & \mbox{if $j>n_2^*(0)$}\\
	\frac{j(1-b)}{N} & \mbox{if $j\le n_2^*(0)$.}\end{array}\right.
\end{align*}
Since $b<1$, it follows that
\[
\pe(0,n_2^*(0)) \ge \pe(0,j)
\]
 This implies that action $n_2^*(0)$ is a best response to action $0$.

Now we show that any profile $(i, j)$ such that $(i, j)\ne (n_1^*, n_2^*(n_1^*))$
and $(i, j)\ne (0, n_2^*(0))$, cannot be a Nash equilibrium. For each profile 
$(i, j)$ that is not an equilibrium, we identify a better response $i'$ to
action $j$, or a better response $j'$ to action $i$.  In the process of proving
that $(n_1^*, n_2^*(n_1^*))$ is a pure Nash equilibrium, for instance, 
we have shown that actions $n_1^*$ and $n_2^*(n_1^*)$
are best responses to each other.  Thus, to show that profile $(i, j)$ is not a
Nash equilibrium, it is sufficient to assume that 
\begin{equation}
	\begin{array}{l}
		i\ne n_1^*,\ \ i\ne 0 \\
	j\ne n_2^*(n_1^*),\ \ j\ne n_2^*(0).\end{array}\label{noteq}
\end{equation} 
For any profile $(i, j)$ such that (\ref{noteq}) holds, we show that 
\begin{enumerate}
	\item if $j\le N-n_1^*$, then action $n_1^*$ is a better response to action $j$ than action $i$;
	\item if $j> N-n_1^*$, then action $0$ is a better response to action $j$ than action $i$.
\end{enumerate}
Indeed, suppose that $j\le N-n_1^*$, we have
\begin{align*}
	\pu(n_1^*, j) &= 1\cdot \indicatorFunction{n_1^*+j\le N}-\frac{n_1^*b}{N} \\
	&= 1-\frac{n_1^*b}{N}\\
	\pu(i, j) &= \indicatorFunction{\pr(E_{\min(i, N-j)}\le\epsilon}
\indicatorFunction{i+j\le N}-\frac{ib}{N} \\
	&\ \ \left\{\begin{array}{ll}
= -\frac{ib}{N} & \mbox{if $i<n_1^*$} \\
\le 1-\frac{ib}{N} & \mbox{if $i>n_1^*$.}\end{array}\right.
\end{align*}
The preceding equations imply that
\[
\pu(n_1^*, j)\ge \pu(i,j)
\]
for $i\ne n_1^*$ and $j\le N-n_1^*$.  Now assume that $j> N-n_1^*$, we have
\begin{align*}
	\pu(0, j) &= 0-\frac{0\cdot b}{N}=0\\
	\pu(i,j) &= \left\{\begin{array}{ll}
		\indicatorFunction{\pr(E_{\min(i, N-j)})\le \epsilon}\cdot 0-\frac{ib}{N} 
		& \mbox{if $i > n_1^*$}\\
		-\frac{ib}{N} & \mbox{if $1 \le i  \le n_1^*$}\end{array}\right. \\
	&= -\frac{ib}{N}.
\end{align*}
From the preceding equations, clearly $\pu(0, j) > \pu(i,j)$. Thus, we have proven that
action $0$ is a better response than action $i$.  

Finally, we prove the social optimality of these two equilibria.
Let $P_{sum}(i, j)$ denote the sum of payoffs of the two players when action
profile $(i, j)$ is used.  That is,
\[
P_{sum}(i, j)=\pu(i, j)+\pe(i, j).
\]
Now
\begin{align*}
P_{sum}(n_1^*,n_2^*(n_1^*)) &= 1-\frac{n_1^*}{N}\cdot b+\frac{n_2^*(n_1^*)}{N}\cdot (1-b) \\
P_{sum}(0,n_2^*(0)) &= \frac{n_2^*(0)}{N}\cdot (1-b).	
\end{align*}
It follows that 
\begin{align}
& P_{sum}(n_1^*,n_2^*(n_1^*))- P_{sum}(0,n_2^*(0)) = \nonumber\\
&\qquad \frac{N-n_1^*b+(n_2^*(n_1^*)-n_2^*(0))(1-b)}{N}.\label{sopt}
\end{align}
Clearly, $n_1^*\le N$.  From \req{case2}, we have
\[
0\le n_2^*(0)-n_2^*(n_1^*)\le N.
\]
Thus, 
\[
P_{sum}(n_1^*,n_2^*(n_1^*))- P_{sum}(0,n_2^*(0))\ge 0.
\]
If $n_1^* < N$ or $n_2^*(0)-n_2^*(n_1^*)< N$, strategy profile $(n_1^*, n_2^*(n_1^*))$
is strictly socially better than profile $(0, n_2^*(0))$.  On the other hand,
if $n_1^*=N$ and $n_2^*(0)-n_2^*(n_1^*)= N$, the two profiles are equally good socially.

We have completed the proof.

\begin{center}
	{\bf Appendix C}
\end{center}
In this appendix we prove \rprop{opt-prop} and then \rthe{optimal-wf}.  

{\bf Proof of \rprop{opt-prop}.}  We first prove part (1) of \rprop{opt-prop}. 
Since $x_i^*=r_i$, we have
\[
\eta_i-(m-1)\lambda/2 \ge r_i.
\]
From \req{eta_j} we have
\[
\frac{1}{m}\left(\sum_{k=1}^m z_k+L\right)-(m-1)\lambda/2 \ge r_i +z_i \ge z_j+r_j.
\]
Hence,
\[
\eta_j-(m-1)\lambda/2 \ge r_j,
\]
which implies that $x_j^*=r_j$.  We have completed the proof of part (1).  The proof
of part (2) is similar and we omit the proof.  We now consider part (3).  Since 
$0<x_i^*<r_i$ and $0<x_j^*<r_j$, it follows that $\mu_i=\omega_i=\mu_j=\omega_j=0$.
Thus,
\begin{align*}
	x_i^*+z_i &= \frac{1}{m}\left(\sum_{k=1}^m z_k +L\right)-z_i-\frac{(m-1)\lambda}{2}+z_i \\
	&= \frac{1}{m}\left(\sum_{k=1}^m z_k +L\right)-z_j-\frac{(m-1)\lambda}{2}+z_j \\
	&=x_j^*+z_j.
\end{align*}
The proof is completed.

{\bf Proof of \rthe{optimal-wf}.}
Now we prove \rthe{optimal-wf}.  We note that if $\sum_{j=1}^m r_j < L$, the feasible
region of the optimization problem in \req{cnp} is empty.  If $\sum_{j=1}^m r_j = L$, 
it is clear that the feasible region contains only one point.  Thus, the solution is 
$x_j=r_j$ for all $1\le j\le m$. Thus, in the proof we assume that $\sum_{j=1}^m r_j > L$.

Suppose that $L \le \tilde m k$.  In this case, the optimal solution of \req{vcnp} is stated in
Claim 1 below.
\begin{itemize}
	\item {\bf Claim 1} If $L \le \tilde m k$, the optimal solution is $x_j=L/k$ 
	for $1\le j\le k$, and $x_j=0$ for $k+1\le j\le |\N|$.
\end{itemize}
On the other hand, if $\tilde m k < L< \sum_{j=1}^m r_j$, the proof needs the following claims.
\begin{itemize}
	\item {\bf Claim 2} The objective function of $P(\bfz, \bfr, m, L)$ and that of 
	$P(\bfz', \bfr', m, L')$ are the same.
	\item {\bf Claim 3} The objective function is convex.  There is a unique optimal point.
	\item {\bf Claim 4} Assume that $L\ge \tilde m k$.  The feasible region 
	$\Omega'$ of $P(\bfz', \bfr', m, L')$ is a subset of
	the feasible region $\Omega$ of $P(\bfz, \bfr, m, L)$.  In addition, the optimal point
	of $P(\bfz, \bfr, m, L)$ is in $\Omega'$.  
	\item {\bf Claim 5} Since $L< \sum_{j=1}^m r_j$ and each call to $P$ strictly decreases $L$, 
	the recursive function terminates in a finite number of calls.
\end{itemize}
If $L > \tilde m k$, the water-filling algorithm in Algorithm \ref{wfa} raises the water
level by $\tilde m$ and call $P(\bfz', \bfr', m, L')$.  From Claim 2, the original objective
function and the new objective function are the same.  From Claim 3, the constrained nonlinear
program has a unique solution.  From Claim 4, this unique solution is in the feasible region
of $P(\bfz', \bfr', m, L')$.  The water-filling algorithm successively reduces the feasible
region.  In this process, $L$ is strictly reduced.  In a finite number of steps, 
the condition $L \le \tilde m k$ will hold and the algorithm will stop.

Now we prove Claims 1, 2, 4.  Claims 3 and 5 are obvious.  We omit their proofs.\\ 
{\bf Proof of Claim 1.}  First, if $L=0$, it is obvious that the feasible region has only one
point and $x_i=0$ for all $i$ gives the optimal solution.  We assume that $0<L\le \tilde m k$.
Let $\bfx$ denote the optimal solution.  If $x_1=0$, by the second 
statement in \rprop{opt-prop} we have $x_i=0$ for all $2\le i\le |\N|$.  It follows
that $\sum_{i=1}^m x_i=0$, which contradicts the assumption that $L>0$.  
If $x_1 =r_1$, then
\[
\frac{1}{m}\left(\sum_{i=1}^m z_i +L\right)-z_1-\frac{m-1}{2}\lambda > r_1.
\]
Since $z_j=z_1$ for all $j=2, 3, \ldots, k$, it follows that
\[
\frac{1}{m}\left(\sum_{i=1}^m z_i +L\right)-z_j-\frac{m-1}{2}\lambda > r_1.
\]
Thus, $x_j>r_1$.  We have
\[
L=\sum_{i=1}^m x_i \ge \sum_{i=1}^k x_i> r_1 k\ge k \tilde m,
\]
which is a contradiction.  Thus, 
\beq{tmp1}
0<x_1<r_1.
\eeq  
Since $z_1=z_2=\ldots=z_k$ and $r_1\le r_2\le \ldots\le r_k$, it follows 
from the definition of $\eta_i$ in \req{eta_j} that $x_1=x_2=\ldots=x_k$.  We now show 
that $x_j=0$ for $k+1\le j\le |\N|$.  If $z_{k+1}-z_k \ge r_1$, \ie $\tilde m=r_1$, we have
\begin{align*}
	&\eta_{k+1} -\frac{m-1}{2}\lambda \\
	&=\frac{1}{m}\left(\sum_{i=1}^m z_i+L\right)-z_1+z_1-z_{k+1}-\frac{m-1}{2}\lambda \\
	&=x_1+z_1-z_{k+1} \\
	&<r_1-(z_{k+1}-z_k)\le 0.	
\end{align*}
Thus, $x_j=0$ for $k+1\le j\le |\N|$.  On the other hand, consider the case 
where $z_{k+1}-z_k< r_1$, \ie $\tilde m=z_{k+1}-z_k$.  Assume that $x_j>0$ for some
$j=k+1, \ldots, |\N|$.  We look for a contradiction.  Since $x_j>0$, we have
\[
\frac{1}{m}\left(\sum_{i=1}^m z_i+L\right)-z_j-\frac{m-1}{2}\lambda >0.
\]
From the preceding inequality and the definition of $\eta_1$ in \req{eta_j}, we have
\[
\eta_1-\frac{m-1}{2}\lambda> z_j-z_1.
\]
This implies that $x_1> z_j-z_1$.  Thus,
\[
L=\sum_{i=1}^m x_i \ge \sum_{i=1}^k x_i > k(z_j - z_1)\ge k \tilde m.
\]
This is a contradiction.

{\bf Proof of Claim 2.}  We assume that $L> \tilde m k$.
We use notation $\eta_j(\bfz, \bfr, L)$ to explicitly show the dependency with
$\bfz, \bfr, $ and $L$.
From \req{eta_j} and $\bfz', \bfr'$, and $L'$ defined in (\ref{z'}), (\ref{r'}) and (\ref{L'}), we have
\begin{align}
	& \eta_j(\bfz', \bfr', L') \\
	&= \frac{1}{m}\left(\sum_{\ell=1}^m z_\ell'(t) +L'\right)-z_j'(t) \nonumber\\
	&= \frac{1}{m}\left(\sum_{\ell=1}^k z_\ell'(t) +\sum_{\ell=k+1}^m z_\ell'(t)  +L'\right)-z_j'(t) \nonumber\\
	&= \frac{1}{m}\left(\sum_{\ell=1}^k (z_\ell(t)+\tilde m) +\sum_{\ell=k+1}^m z_\ell(t)  +
	L-\tilde m k\right) \nonumber\\
     &\quad-z_j'(t) \nonumber\\
	&= \eta_j(\bfz, \bfr, L)- \tilde m \indicatorFunction{1\le j\le k},\label{Aj-equiv}
\end{align}
where $\indicatorFunction{A}$ is equal to $1$ if $A$ is true, and is equal to $0$ otherwise.
It follows from (\ref{Aj-equiv}) that one can rewrite 
\[
x_j-\eta_j(\bfz, \bfr, L)=y_j-\eta_j(\bfz', \bfr', L')
\]
through a change of variables
\[
y_j=x_j - \tilde m \indicatorFunction{1\le j\le k}.
\]
Thus, the objective functions in $P(\bfz, \bfr, m, L)$ and $P(\bfz', \bfr', m, L')$ are identical.

{\bf Proof of Claim 4.} Let $\bfx$ and $\bfy$ denote the solution of problems 
$P(\bfx, \bfr, m, L)$ and $P(\bfz', \bfr', m, L')$, respectively, where $\bfx$ and
$\bfy$ are related by \req{opt-x-1} or \req{opt-x-2} depending on whether
$r_1<z_{k+1}-z_k$ or $r_1\ge z_{k+1}-z_k$.  We assume that indexes are labeled such
that \req{ordering-1} and \req{ordering-2} hold.  Recall that $\bfz'$, $\bfr'$ 
and $L'$ are defined in (\ref{z'}), (\ref{r'}) and (\ref{L'}), respectively.
The feasible region of problem $P(\bfz, \bfr, m, L)$ is
\beq{original-omega}
\Omega=\left\{\bfx : \sum_{i=1}^m x_i=L, 0\le x_i\le r_i, 1\le i\le m\right\}.
\eeq
In terms of $\bfy$, feasible region $\Omega'$ is
\begin{align*}
\Omega' &=\Biggl\{\bfy : \sum_{i=1}^m y_i=L', 
0\le y_i\le r_i', 1\le i\le m\Biggr\}.
\end{align*}
If $r_1<z_{k+1}-z_k$, one can apply a change of variables using \req{opt-x-1}
to express $\Omega'$ as
\begin{align*}
	\Omega' &=\Biggl\{\bfx : \sum_{i=1}^j 0 +\sum_{i=j+1}^k (x_i-\tilde m)+\sum_{i=k+1}^m x_i=L',\\ 
	&\quad x_i=\tilde m, 1\le i\le j, \\
	&\quad 0\le x_\ell -\tilde m\le r_\ell-\tilde m,
	j+1\le \ell \le k, \\
	&\quad 0\le x_\nu\le r_\nu, k+1\le\nu\le m\Biggr\}.
\end{align*}
With a little manipulation, we obtain from the preceding
\begin{align}
	\Omega' &=\Biggl\{\bfx : \sum_{i=1}^m x_i=L, x_i=r_i, 1\le i\le j, \nonumber\\
	&\quad 0\le x_\ell\le r_\ell, j+1\le\ell\le m\Biggr\}.\label{new-omega}
\end{align}
From \req{original-omega} and (\ref{new-omega}), it is clear that $\Omega'\subseteq \Omega$.
The proof for the case, where $r_1\ge z_{k+1}-z_k$ is similar.  We omit the details.

We now show that the optimal point of problem $P(\bfz, \bfr, m, L)$ actually lies in
region $\Omega'$.  We prove the case where 
\beq{tmp5}
r_1 < z_{k+1}-z_k.
\eeq  
From (\ref{new-omega}), we need to show that the optimal solution 
satisfies $x_i=r_i$ for $i=1, 2, \ldots, j$.  Assume that $x_1 < r_1$.  
From \req{KKT-1}, since $x_1<r_1$, it follows that $\mu_1=\omega_1=0$ and
\beq{tmp3}
\frac{1}{m}\left(\sum_{\ell=1}^m z_\ell +L\right)-z_1-\frac{m-1}{2}\lambda <r_1.
\eeq
Since $z_1=z_i$ for $i=2, 3, \ldots, k$, it follows that
\[
x_i=x_1<r_1.
\]
In addition, for $i=k+1, \ldots, m$, we have
\begin{align}
&\frac{1}{m}\left(\sum_{\ell=1}^m z_\ell +L\right)-z_i-\frac{m-1}{2}\lambda \nonumber \\
&=\frac{1}{m}\left(\sum_{\ell=1}^m z_\ell +L\right)-z_1-\frac{m-1}{2}\lambda +z_1-z_i.
\label{tmp4}
\end{align}
From \req{tmp3} and (\ref{tmp4}), we have
\begin{align*}
&\frac{1}{m}\left(\sum_{\ell=1}^m z_\ell +L\right)-z_i-\frac{m-1}{2}\lambda  \\
&< r_1+z_1-z_i \\
&\le 0,	
\end{align*}
where the last inequality is due to the assumption \req{tmp5}.
Thus, $x_i=0$ for $i=k+1, \ldots, m$.  It follows that
\[
L=\sum_{\ell=1}^m x_\ell = \sum_{\ell=1}^k x_\ell+ \sum_{\ell=k+1}^m x_\ell< \tilde m k,
\]
which contradicts the assumption that $L>\tilde m k$. The proof for the other case where
$r_1 \ge z_{k+1}-z_k$ is similar.  We omit the details.
The proof is complete.

\bigskip
% To print the credit authorship contribution details

\bibliography{bibdatabase}
\bibliographystyle{ieeetran}

\bigskip
\newpage
\end{document}